\documentclass[twocolumn,epjc3]{svjour3} 

\newcommand{\begg}{\begin{gather}}
\newcommand{\beq}{\begin{equation}}

\newcommand{ \ds}{\displaystyle}
\newcommand{\eegg}{\end{gather}}
\newcommand{\eeq}{\end{equation}}

\newcommand{\Rho}{\mathrm{P}}

\RequirePackage[colorlinks,citecolor=blue,urlcolor=blue,linkcolor=blue]{hyperref}
\RequirePackage{flushend}
\RequirePackage{mathptmx} 
\RequirePackage[numbers,sort&compress]{natbib}
\RequirePackage[T1]{fontenc}

\smartqed 

\usepackage{amsmath}
\usepackage{amssymb}
\usepackage{citesort}
\usepackage{color}
\usepackage{dcolumn}
\usepackage{graphicx}
\usepackage{hyperref}
\usepackage{url}
\usepackage[utf8]{inputenc}
\usepackage{booktabs,makecell,multirow,tabularx}

\journalname{Eur. Phys. J. C}

\begin{document}
\newpage

\title{Photon sector analysis of Super and Lorentz symmetry breaking: effective photon mass, bi-refringence and dissipation}

\author{\mbox
Luca Bonetti\thanksref{e1,addr1,addr2}
        \and
        Lu{\'i}s R. dos Santos Filho\thanksref{e2,addr3}
        \and
        Jos{\'e} A. Helay{\"e}l-Neto\thanksref{e3,addr3} 
        \and \\
        Alessandro D.A.M. Spallicci\thanksref{e4,addr1,addr2,addr3}
}

\thankstext{e1}{e-mail: luca.bonetti696@gmail.com}
\thankstext{e2}{e-mail: luis.um.dia.seja@gmail.com}
\thankstext{e3}{e-mail: josehelayel@gmail.com}
\thankstext{e4}{e-mail: spallicci@cnrs-orleans.fr\\
Web page: http://wwwperso.lpc2e.cnrs.fr/$\sim$spallicci/\\
Corresponding author}

\institute{Universit\'e d'Orl\'eans\\
\mbox{Observatoire des Sciences de l'Univers en r\'egion Centre (OSUC) UMS 3116} \\
\mbox{1A rue de la F\'{e}rollerie, 45071 Orl\'{e}ans, France}\\
\mbox{Collegium Sciences et Techniques (COST), P\^ole de Physique}\\
\mbox{Rue de Chartres, 45100  Orl\'{e}ans, France}\label{addr1}
          \and
          Centre Nationale de la Recherche Scientifique (CNRS)\\
\mbox{Laboratoire de Physique et Chimie de l'Environnement et de l'Espace (LPC2E) UMR 7328}\\
\mbox {Campus CNRS, 3A Av. de la Recherche Scientifique, 45071 Orl\'eans, France}\label{addr2}
          \and
          Centro Brasileiro de Pesquisas F\'{\i}sicas (CBPF)\\
\mbox{Departamento de Astrof\'{\i}sica, Cosmologia e Intera\c{c}\~{o}es Fundamentais (COSMO)}
\mbox {Rua Xavier Sigaud 150, 22290-180 Urca, Rio de Janeiro, RJ, Brasil}\label{addr3}
}

\date{File BlDlHjSa-EPJC-180901-1-as.tex / 
Accepted: 16 September 2018}

\maketitle

\begin{abstract}
Within the Standard Model Extension (SME), we expand our previous findings on four classes of violations of Super-Symmetry (SuSy) and Lorentz Symmetry (LoSy), differing in the handedness of the Charge conjuga\-tion-Parity-Time reversal  
(CPT) symmetry and in whether considering the impact of photinos on photon propagation. The violations, occurring at the early universe high energies, show visible traces at present in the Dispersion Relations (DRs). For the CPT-odd classes ($V_{\mu}$ breaking vector) associated with the Car\-roll-Field-Jackiw (CFJ) model, the DRs and the Lagrangian show for the photon an effective mass, gauge invariant, proportional to 
$|{\vec V}|$. The group velocity exhibits a classic dependency on the inverse of the frequency squ\-ared.  
For the CPT-even classes ($k_{F}$ breaking tensor), when the photino is considered, the DRs display also a massive behaviour inversely proportional to a coefficient in the Lagrangian and to a term linearly dependent on $k_{F}$. All DRs display  an angular dependence and lack LoSy invariance.
In describing our results, we also point out the following properties: i) the appearance of complex
or simply imaginary frequencies and super-luminal speeds and ii) the emergence of bi-refringence.
Finally, we point out the circumstances for which SuSy and LoSy breakings, possibly in presence of
an external field, lead to the non-conservation of the photon energy-momentum tensor. We do so
for both CPT sectors.
\end{abstract}


\section{Introduction, motivation and structure of the work}


For the most part, we base our understanding of particle physics on the Standard Model (SM). The SM proposes the Lagrangian of particle physics
and summarises three interactions among fundamental particles, accounting for electromagnetic (EM), weak and strong
nuclear forces. The mo\-del has been completed theoretically in the mid seventies, and has found several experimental confirmations
ever since. In 1995, the top quark was found \cite{Abetal95}; in 2000, the tau neutrino was directly measured \cite{koetal00}. Last,
but not least, in 2012 the most elusive particle, the Higgs Boson, was found \cite{Aad2015}. The associated Higgs field induces the spontaneous symmetry breaking mechanism, responsible for all the masses of the SM particles. Neutrinos and the photon remain massless, for they do not have a direct interaction with the Higgs field. Remarkably, massive neutrinos are not accounted for by the SM.

All ordinary hadronic and leptonic matter is made of Fermions, while Bosons are the
interaction carriers in the SM. The force carrier for the electromagnetism is the photon. Strong nuclear interactions are mediated by eight
gluons, massless but not free particles, described by Quantum Chromo-Dynamics (QCD). Instead, the $W^{+}$, $W^{-}$ and $Z$ massive Bosons,
are the mediators of the weak interaction. The charge of the W-mediators has suggested that the EM and weak nuclear forces can be unified into a single interaction called electroweak interaction. 

We finally notice that the photon is the only massless non-confined Boson; the reason for this must at least be questioned by fundamental physics.

SM considers all particles being massless, before the Higgs
field intervenes. Of course, masslessness of particles would be in contrast with every day experience. In 1964, Higgs
and others \cite{higgs1964,englertbrout1964,guralnikhagenkibble1964} came up with a mechanism that, thanks to the introduction of a new field - the
Higgs field - is able to explain why the elementary particles in the spectrum of the SM, namely, the charged leptons and quarks, become massive. But the detected mass of the Higgs
Boson is too light: in 2015 the ATLAS and CMS experiments showed that the Higgs Boson mass is $125.09\pm0.32\ $ GeV/c$^2$ \cite{Aad2015}.
Between the GeV scale of the electroweak interactions and the Grand Unification Theory (GUT) scale ($10^{16}$ GeV), it is widely believed that new physics should appear at the TeV scale, which is now the experimental limit up to which the SM was tested \cite{Lykken2010}. Consequently, we need a fundamental theory that reproduces the phenomenology at the electroweak scale and, at the same time, accounts for effects beyond the TeV scale. 

An interesting attempt to go beyond the SM is for sure Super-Symmetry (SuSy); see \cite{martin2016} for a review. This theory predicts the existence of new particles that are not included in the SM. The interaction between the Higgs and these new SuSy particles would cancel out some SM contributions to the Higgs Boson mass, ensuring its lightness. This is the solution to the so-called gauge hierarchy problem. The SM is assumed to be Lorentz\footnote{Usually, the Lorentz transformations describe rotations in space (J symmetry) and boosts (K symmetry) connecting uniformly moving bodies. When they are complemented by translations in space and time (symmetry P), the transformations include the name of Poincar\'e.} Symmetry (LoSy) invariant. Anyway, it is reasonable to expect that this prediction is valid only up to certain energy scales \cite{kosteleckysamuel1989a,kosteleckysamuel1989b,kosteleckysamuel1989c,kosteleckysamuel1991,kosteleckypotting1991,
kosteleckypotting1995,kosteleckypotting1996},
beyond which a LoSy Violation (LSV) might occur. The LSV would take place following  the condensation of tensor fields in the context of open Bosonic strings.

The aforementioned facts show that there are valid reasons to undertake an investigation of physics beyond the SM and also
consider LSV. There is a general framework where we can test the low-energy manifestations of LSV, the so-called Standard Model Extension (SME) \cite{colladaykostelecky1997,colladaykostelecky1998,myerspospelov2003,liberati2013}. Its effective Lagrangian is  
given by the usual SM Lagrangian, modified by a combination of SM operators of any dimensionality contracted with Lorentz breaking tensors of suitable rank to get a scalar expression for the Lagrangian. 

For the Charge conjugation-Parity-Time reversal (CPT) odd classes the breaking factor is the $V_{\mu}$ vector associated with the Carroll-Field-Jackiw (CFJ) model \cite{cafija90}, while for the CPT-even classes it is the $k_{F}$  tensor. 

In this context, LSV has been thoroughly investigated phenomenologically. Studies include electron, photon, mu\-on, meson,
baryon, neutrino and Higgs sectors \cite{kosteleckyrussell2011}. Limits on the parameters associated to the breaking of relativistic
covariance are set by quite a few experiments \cite{kosteleckyrussell2011,kosteleckytasson2009,baileykostelecky2004}. LSV has also been tested
in the context of EM cavities and optical systems \cite{russell2005,phhumastvewa2001,bestwakola2000,huphmavest2003,mubrhepela2003,muhesapela2003,muller2005}.
Also Fermionic models in presence of LSV have been proposed: spinless and/or neutral particles with a non-minimal coupling
to a LSV background, magnetic properties in relation to Fer\-mionic matter or gauge Bosons 
\cite{
besifeor2011,
bakkebelichsilva2011a,
bakkebelichsilva2011b,
bakkebelich2012,
casanaferreirasantos2008a,
casanaferreirasantos2008b,
casanaferreiragomespinheiro2009a,
casanaferreiragomespinheiro2009b,
casanaferreirarodriguessilva2009,
limabelichbakke2013,
cvdshobe2014,
hernaskibelich2014}.
 
   More recently, \cite{fu-lehnert-2016,colladay-noordmans-potting-2017} present interesting results involving the electroweak sector of the SME.

\begin{figure}
\includegraphics[width=9cm,height=6cm]{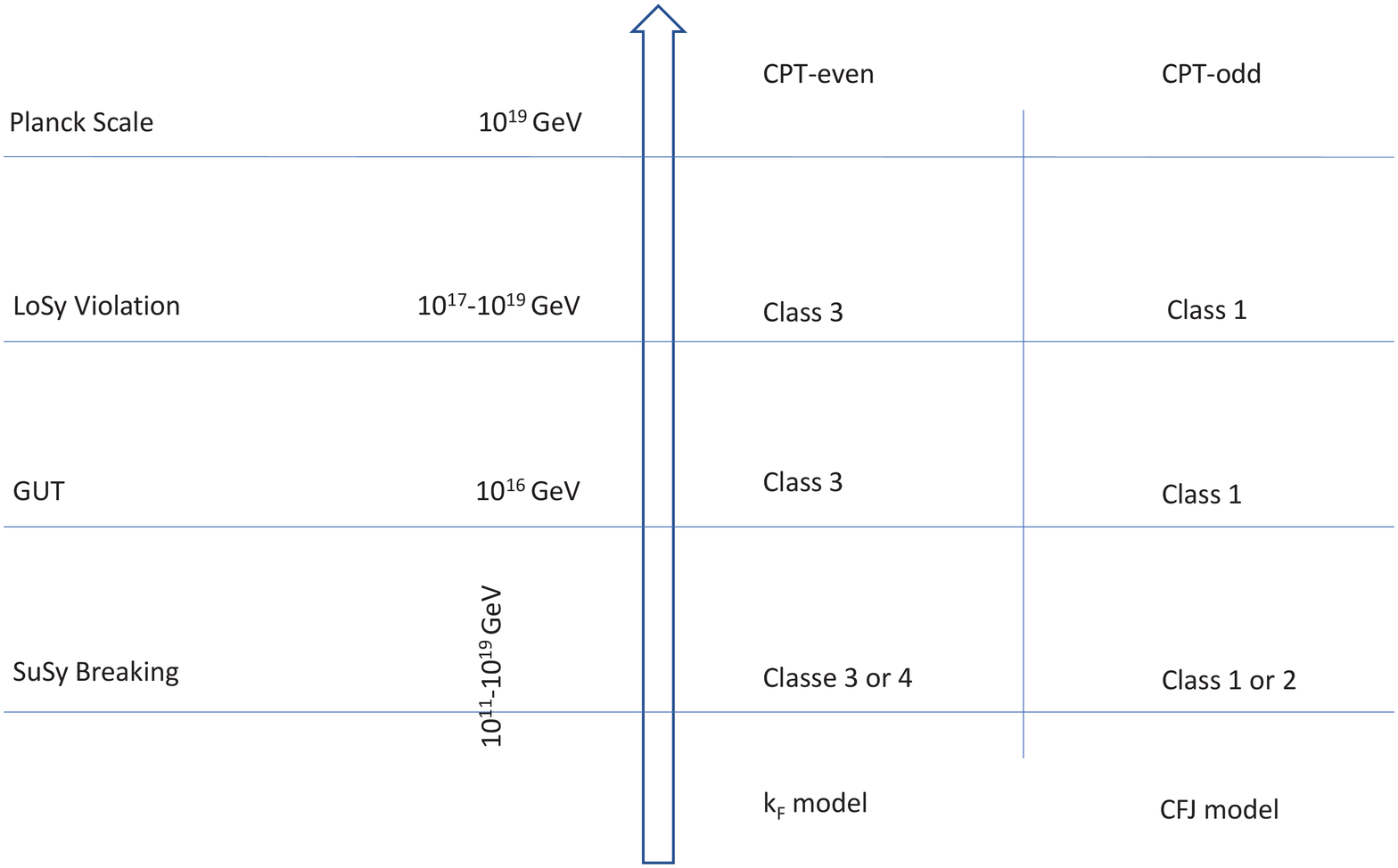}
\caption{We show the energy scales at which the symmetries are supposed to break, referring to the model described in \cite{bebegahnle2015}. At Planck scale, $10^{19}$ GeV, all symmetries are exact, unless LoSy breaking occurs. This latter may intervene at a lower scale of $10^{17}$ GeV, but anyway above GUT. Between $10^{11}$ and $10^{19}$ GeV, we place the breaking of SuSy. In our analysis, we assume that the four cases of SuSy breaking occur only when LoSy has already being violated. Interestingly, at our energy levels, we can detect the reminiscences of these symmetry breakings.}  
\label{fig1}
\end{figure}

Following 
\cite{
colladaymcdonald2011,
bergerkostelecky2002,
falenape2012,
bolokhovnibbelinkpospelov2005,
katzshadmi2006,
nibbelinkpospelov2005,
redigolo2012,
golenapeds2013,
lenapeds2013,
hnbedilesp2010},
LSV is stemmed from a more fundamental physics because it concerns higher energy levels of those obtained in particle accelerators. In Fig. \ref{fig1}, we show the energy scales at which the symmetries are supposed to break, referring to the model described in  
\cite{bebegahnle2015}. At Planck scale, $10^{19}$ GeV, all symmetries are exact, unless LoSy breaking occurs. This latter may intervene at a lower scale of $10^{17}$ GeV, but anyway above GUT. Between $10^{11}$ and $10^{19}$ GeV, we place the breaking of SuSy. In our analysis, we assume that the four cases of SuSy breaking occur only when LoSy has already being violated. Interestingly, at our energy levels, we can detect the reminiscences of these symmetry breakings. 

Indeed, we adopt the point of view that the LSV background is part of a SuSy multiplet; see for instance
 \cite{bebegahnle2015}.

Since gravitational wave astronomy is at its infancy, EM wave astronomy remains the main detecting tool for unveiling the universe. Thereby, testing the properties of the photons is essential to fundamental physics and astrophysics has just to interpret the universe accordingly.

A legitimate question addresses which mechanism could provide mass to the photon and thereby how the SM should be extended to accommodate such a conjecture.
We have set up a possible scenario to reply to these two questions with a single answer.

Non-Maxwellian massive photon theories have been proposed over the course of the last century. If the photon is massive, propagation is affected in terms of group velocity and polarisation. 

This work is structured as follows. In Sect. 2, 
we summarise, complement and detail the results obtained in our letter \cite{bodshnsp2017}, with some reminders to the appendix. Within the unique SME model, we consider four class\-es of models that exhibit LoSy and SuSy violations, varying in CPT handedness and in incorporating - or not - the effect of photino on the photon propagation. The violation occurs at very high energies, but we search for traces in the DRs visible at our energy scales.  
In the same Section, we confirm that  a massive photon term emerges from the CPT-odd Lagrangian. We discover that a massive photon emerges also for the CPT-even sector when the photino is considered. We also point out when i) complex or simply imaginary frequencies and super-luminal speeds arise.  In Sect. 3, we look for multi-fringence. In Sect. 4, we wonder if dissipation is conceivable for wave propagation in vacuum and find an affirmative answer. In Sect. 5, we propose our conclusions, discussion and perspectives. The appendix gives some auxiliary technical details. 


\subsection{Reminders and conventions}


We shall encounter real frequencies sub- and luminal velocities but also imaginary and complex frequencies, and super-luminal velocities\footnote{A velocity $v$ larger than $c$ is associated to the concept of tachyon \cite{recami1986,recamifontanagaravaglia2000} and implies an imaginary relativistic factor $\gamma$. If wishing (relativistic) energy $E$ and (relativistic) mass $m$ to remain real, rest mass $m_0$ must be imaginary
\beq
E = mc^2 = \gamma m_0 c^2 = \frac{m_0c^2}{\sqrt{ 1 - {\ds\frac{v^2}{c^2}}}}~.
\eeq
Similarly, wishing measured frequency $f$ to remain real, frequency $f_0$ must be imaginary in the rest frame
\beq
f = \frac{f_0}{\gamma} = \nu_0 \sqrt{ 1 - {\ds\frac{v^2}{c^2}}}~.
\eeq
Alternatively, letting rest mass and rest frequency real, mass and energy become imaginary. In the particle view, recalling that 
$E= h \nu$, we recover both interpretations. 
An imaginary frequency implies an evanescent wave amplitude, and thereby tachyonic modes are associated to 
transitoriness. 
Complex frequencies present the features above for the imaginary part, and usual properties for the real part.
Finally, few scholars consider causality not necessarily incompatible with tachyons 
\cite{aharonovkomarsusskind1969,recami1977,hawking1985,garrison-etal-1998,hawkinghertog2002,liberati-sonego-visser-2002,shabad-usov-2011,schwartz2016}. }.

In this work, see the title, we intend photon mass as an effective mass. The photon is dressed of an effective mass, that we shall see, depends on the perturbation vector or tensor. Nevertheless,  we are cautious in differentiating an effective from a real mass. The Higgs mechanism gives masses to the charged leptons and quarks, the W and Z bosons, while the composite hadrons (baryons and mesons), built up from the massive quarks, have most of their masses from the mechanism of Chiral Symmetry (Dynamical) Breaking (CSB). It would be epistemologically legitimate  to consider such mechanisms as producing an effective mass to particles which, without such dressing mechanisms, would be otherwise massless. What is then real or effective? The feature of being frame dependent renders surely the concept of mass unusual, but still acceptable to our eyes, being the dimension indeed that of a mass. 

We adopt natural units for which $c=\hbar=1/4\pi\epsilon_0 = \mu = 1$, unless otherwise stated. We adopt the metric signature as \-  
(+, -, -, -). 
Although more recent literature adopts $k_{AF}^\mu$ and $k_{F}^{\mu\nu\rho\sigma}$ for LSV vector and tensor, respectively, we drop the former in favour of $V^\mu$ for simplicity of notation especially when addressing time or space components and normalised units. 

Finally, we omit to use the adjective angular, when addressing the angular frequency $\omega$.  

\subsection{ Upper limits on $V_{\mu}$ vector and photon mass $m_\gamma$\label{upperlim}}

Ground based experiments indicate that $|{\vec V}|$, the space components, must be smaller than $10^{-10}$ eV $ = 1.6 \times 10^{-29}$ J from the bounds given by the 
energy shifts in the spectrum of the hydrogen atom 
\cite{gomesmalta2016}; else smaller than $8\times 10^{-14}$ eV $ = 1,3 \times 10^{-32}$ J from measurements of the rotation in the polarisation of light in
resonant cavities \cite{gomesmalta2016}.
The time component of $V_\mu$ is smaller than $10^{-16}$ eV $ = 1.6 \times 10^{-35}$ J \cite{gomesmalta2016}
 Instead, astrophysical observations lead to  
 $|{\vec V}|< 10^{-34}$ eV $ = 1.6 \times 10^{-53}$ J. We cannot refrain to remark that such estimate is equivalent to the Heisenberg limit ($\Delta m\Delta t >1$) on the smallest measurable energy or mass for a given time t, set equal to the Universe age.  
The actual Particle Data Group (PDG) limit on photon mass  \cite{tanabashietal2018} refers to values obtained in \cite{Ryutov1997,Ryutov2007} of $10^{-54}$ kg or $5.6 \times 10^{-19}$ eV/c$^2$, to be taken with some care, as motivated in \cite{retinospalliccivaivads2016,boelmasasgsp2016,boelmasasgsp2017}.



\section{ LSV and two classes of SuSy breaking for each CPT sector}


We summarise and complement in this section the results obtained in \cite{bodshnsp2017}.


\subsection{CPT-odd sector and the $V_{\mu}$ vector: classes 1 and 2}


The CFJ proposition \cite{cafija90} introduced LSV by means of a Chern-Simons (CS) \cite{chernsimons1974} term in the Lagrangian that represents the EM interaction. It was conceived and developed outside any SuSy scenario. The works \cite{baetaetal2004}  
and later \cite{bebegahnle2015} framed the CFJ model in a SuSy scenario. The LSV is obtained through the breaking vector $V_{\mu}$,  
the observational limits of which are considered in the CFJ framework. For the origin, the microscopic justification was traced in the fundamental Fermionic condesates present in SuSy \cite{bebegahnle2015}. In other words, the Fermionic fields present in the in SuSy background may condensate (that is, take a vacuum expectation value), thereby inducing LSV.

In the following, the implications of the CS term on the propagation and DR of the photon are presented. 


\subsubsection{Class 1: CFJ model}


The Lagrangian reads

\beq
L_{1}=-\frac{1}{4}F^{\mu\nu}F_{\mu\nu}-\frac{1}{2}\epsilon^{\mu\nu\sigma\rho}V_{\mu}A_{\nu}F_{\sigma\rho}~.
\label{CFJ_lagrangian}
\eeq
where $F_{\mu\nu} = \partial_\mu A_\nu - \partial_\nu A_\mu$ and $F^{\mu\nu} = \partial^\mu A^\nu - \partial^\nu A^\mu$ are the covariant and contravariant forms, respectively, of the EM tensor; $\epsilon^{\mu\nu\sigma\rho}$ is the contravariant form of the Levi-Civita pseudo-tensor, and $A_\mu$ the potential covariant four-vector.   

We observe the coupling between the EM field and the breaking vector $V_{\mu}$. The Euler-Lagrange variational principle applied
to Eq. (\ref{CFJ_lagrangian}) leads to 

\beq
\vec{\nabla}\times \vec{B}+V_{0}\vec{B}-\vec{V}\times\vec{E}=\partial_{t}\vec{E}~. 
\label{modeb}
\eeq 
where the three-vector $\vec V$ represents the space components of $V_{\mu}$, and $\vec{B}$ and $\vec{E}$ the magnetic and electric fields. 

From the Fourier transformation of the curl of the electric field $(\vec{\nabla} \times \vec{E}) $ equation, we obtain 
$\tilde{\vec{B}}$ in terms of $\tilde{\vec{E}}$, magnetic and electric field in Fourier domain, respectively

\beq
\vec{\tilde{B}}={ \frac{\vec {k}}{\omega}}\times\vec{\tilde{E}}~,
\label{vectildeb}
\eeq
where the four-momentum is $k^{\mu}=\left(\omega,\vec{k}\right)$ and where $k^2 = (\omega^2 - {\bf k^2})$. 
Inserting Eq. (\ref{vectildeb}) into the Fourier transform of Eq. (\ref{modeb}), we get
\begin{gather}
\left(\omega^{2}-\vec{k}^{2}\right)\vec{\tilde{E}}+\left(\vec{k}\cdot\vec{\tilde{E}}\right)\vec{k}=i\left(V_{0}\vec{k}\times\vec{\tilde{E}}-\omega\vec{V}\times\vec{\tilde{E}}\right)~.\label{CFJ_DR}
\end{gather}

Equation (\ref{CFJ_DR}) can be arranged in the form
\beq
R_{ij}\tilde{E}_{j}=0~,
\eeq
where $R_{ij}$ is the matrix
\beq
R_{ij}=i\vec{k}^{2}\delta_{ij}+ik_{i}k_{j}-V_{0}\epsilon_{ijk}k_{k}+\epsilon_{ijk}\omega V_{k}~.
\eeq

Imposing $\mbox{det}~R_{ij}=0$, we derive the DR, Eq. (3) in \cite{bodshnsp2017}, known since the appearance of \cite{cafija90}

\beq
\left(k^{\mu}k_{\mu}\right)^{2}+\left(V^{\mu}V_{\mu}\right)\left(k^{\nu}k_{\nu}\right)-\left(V^{\mu}k_{\mu}\right)^{2}=0~.
\label{CFJ DR}
\eeq


\subsubsection{Class 2: Supersymmetrised CFJ model and SuSy breaking}


We can study the effect of the photino on the photon propagation. 
For accounting for the
effects of the photino, according to \cite{bebegahnle2015}, we have to work with the Lagrangian
that follows below 

\begin{equation}
L_{2}=-\frac{1}{4}F +\frac{1}{4}\epsilon^{\mu\nu\rho\sigma}V_{\mu}A_{\nu}F_{\rho\sigma}+\frac{1}{4}HF+M_{\mu\nu}F^{\mu\lambda}{ F^{\nu}_{\lambda}}~,
\label{Lagrangian2}
\end{equation}

where $F = F_{\mu\nu}F^{\mu\nu}$; furthermore, $H$ is a scalar defined in \cite{bebegahnle2015}, the tensor $M_{\mu\nu}=\hat{M}_{\mu\nu}+{\ds \frac{1}{4}}~\eta_{\mu\nu}M$, and $\hat{M}_{\mu\nu}$ depends on the background Fermionic
condensate, originated by SuSy; $\hat M_{\mu\nu}$ is traceless, $M$ the trace of $M_{\mu\nu}$, and $\eta_{\mu\nu}$ the Minkowski metric. The Lagrangian, Eq. (\ref{Lagrangian2}), is rewritten as \cite{bebegahnle2015}
\beq
L_2=-\frac{1}{4}\left(1-H-M\right)F+\frac{1}{4}\epsilon^{\mu\nu\rho\sigma}V_{\mu}A_{\nu}F_{\rho\sigma}+\hat{M}_{\mu\nu}F^{\mu\lambda}F_{\lambda}^{\nu}~.
\eeq

In \cite{bebegahn2013} 
it is shown that the DR is equivalent to Eq. (\ref{CFJ DR}), but for a rescaling of the breaking vector. The latter is obtained by integrating out the Fermionic SuSy partner, the photino. The following DR comes out (Eq. (6) in \cite{bodshnsp2017})

\beq
\left(k^{\mu}k_{\mu}\right)^{2}+\frac{\left(V^{\mu}V_{\mu}\right)\left(k^{\nu}k_{\nu}\right)}{\left(1-H-M\right)^{2}}-\frac{\left(V^{\mu}k_{\mu}\right)^{2}}{\left(1-H-M\right)^{2}}=0~.
\label{DR-class2}
\eeq

The background parameters are very small, being suppressed by powers of the Planck energy; they render the denominator in 
Eq. (\ref{DR-class2}) close to unity, implying similar numerical outcomes for the two dispersion relations of Classes 1 and 2.  Consequently, we shall derive and work with group velocities and time delays, for Classes 1 and 2 in a single treatment.


\subsubsection{Group velocities and time delays for Classes 1 and 2}


\paragraph{Zero time component of the breaking vector.\label{221}} \mbox{} \\


We pose $V_{0} = 0$ and rewrite Eq. (\ref{CFJ DR}) as

\beq 
\omega^{4}- { {\cal A}}\omega^{2} + { {\cal B}} =0~, 
\label{omega4ab}
\eeq

having defined 

\[
{ {\cal A}} =2|\vec{k}|^{2}+|\vec{V}|^{2}~~~~~~~~~~
{ {\cal B}} = |\vec{k}|^{4}+|\vec{k}|^{2}|\vec{V}|^{2}-\left(\vec{V}\cdot\vec{k}\right)^{2}~.
\]

The dispersion relation yields 

\beq
\omega^2 - |{\vec k}|^2 = k^\mu k_\mu = \frac{|{\vec V}|^2}{2} + { p} |{\vec V}|\left (\frac{|{\vec V}|^2}{4} + |{\vec k}|^2 \cos^2\theta~\right)^{1/2}~,
\label{drabalpha}
\eeq  
where ${ p} = \pm 1$ and $\theta$ is the angle between ${\vec V}$ and ${\vec k}$. 

For ${ p} = -1$ and $\cos\theta \neq 0$, we get $k^\mu k_\mu < 0$, that is $k_\mu$ space-like and tachyonic velocities. Still for ${ p}= -1$, but  
$\cos\theta = 0$, that is the wave propagating orthogonally to $\vec V$, we obtain $\omega^2 =  |{\vec k}|^2$ and thus a Maxwellian propagation, luxonic velocities, in this specific direction.  

Instead, ${ p} = 1$ leads to $k^\mu k_\mu = m_\gamma^2$, that is $k_\mu$ time-like and bradyonic velocities associated to a massive photon. 

Specifically in the massive photon rest frame, $\vec k = 0$, we get $m_{\gamma}^2 = |{\vec V}|^2$.
Rearranging Eq. (\ref{omega4ab},) we get $|{\vec k}|$ in terms of $\omega^2$   

\beq
|{\vec k}|^2 - \omega^2 = - \frac{1}{2} |{\vec V}|^2 \sin^2\theta  \pm |{\vec V}|\left (\frac{|{\vec V}|^2}{4} + \omega^2 \cos^2\theta\right)^{1/2}~.
\label{omega4abarr1}
\eeq 

Now the plus sign yields $\omega^2 - |{\vec k}|^2 = k^\mu k_\mu <0$, whereas the minus sign is compatible with causal propagation. We rewrite Eq. (\ref{omega4abarr1}) as  

\beq
\frac{|{\vec k}|^2} {\omega^2} =  1 - \frac{|{\vec V}|^2}{2\omega^2} \sin^2\theta + 
{ q} \left (\frac{|{\vec V}|^4}{4\omega^4}\sin^4\theta + \frac{|{\vec V}|^2}{\omega^2}\cos^2\theta \right)^{1/2},
\label{omega4abarr2}
\eeq 
with ${ q} = \pm 1$. If ${ q} = 1$, we recover the case associated with ${ p} = -1$, while for ${ q}=-1$ the case associated with ${ p}=1$. Given the anisotropy introduced by $|{\vec V}|$, we no longer identify the group velocity as 

\beq 
v_g = \frac{\partial \omega}{\partial |{\vec k}|}~, 
\eeq

and instead compute the components of $v_g$

\beq 
v_{gi} = \frac{\partial \omega}{\partial k_i}~,  
\eeq

and thereby have 
\beq 
|\vec{v}_g|^2 = v_{gi}v_{gi}~.
\eeq
having used summation on the $i$ index.
Deriving Eq. (\ref{omega4ab}) with respect to $k_i$, we get 

\beq
v_{gi} = \frac{k_i}{\omega} + \frac{{\vec V} \cdot {\vec k}}{2\omega^2 - 2|{\vec k}|^2 - |{\vec V}|^2} \frac{V_i}{\omega}~,
\eeq

and using Eq. (\ref{drabalpha}), we are finally able to write  

\beq
{\vec v}_g = \frac{\vec k}{\omega} + { p} \frac{|{\vec k}|} {\omega} \frac {\cos \theta}{(|{\vec V}|^2 + 4 |{\vec k}|^2 \cos^2\theta)^{1/2}} 
{\vec V}~, 
\eeq

and 

\begin{eqnarray}
|{\vec v}_g| =  \frac{|{\vec k}|}{\omega}& & \left[ 1+ 2{ p} |{\vec V}|\frac{\cos^2 \theta} {(|{\vec V}|^2 + 4 |{\vec k}|^2 \cos^2\theta)^{1/2}}+ \right. \nonumber \\
& & \left. |{\vec V}|^2 \frac{\cos^2 \theta} {|{\vec V}|^2 + 4 |{\vec k}|^2 \cos^2\theta} \right ]^{1/2}~. 
\label{modvg}
\end{eqnarray}

Through Eq. (\ref{omega4abarr2}), and recalling the conditions ${ p} = 1$ or ${ q} = -1$ for $k_\mu$ time-likeness ($k^2 >0$), Eq. (\ref{modvg}) may be cast as function of $\omega^2$. We consider special cases, starting with $\cos \theta=0$ and have after some computation 

\beq
|{\vec v}_g| = \left [1 - \left(\frac{|{\vec V}|}{\omega}\right)^2\right]^{1/2} =  
1 - \frac{1}{2} \left(\frac{|{\vec V}|}{\omega}\right)^2 + \mathcal {O} \left(\frac{|{\vec V}|}{\omega}\right)^4 ~,  
\label{modvgcostheta0}
\eeq
while for a parallel or anti-parallel propagation to the LSV vector, we get      
\beq
|{\vec v}_g| = 
1 - \frac{1}{8} \left(\frac{|{\vec V}|}{\omega}\right)^2 + \mathcal {O} \left(\frac{|{\vec V}|}{\omega}\right)^4~. 
\label{modvgcostheta1}
\eeq

If we consider { experiment} based limits on $|{\vec V}|$, see Sect. \ref{timedelays}, they determine that the ratio $|{\vec V}|/\omega$ is around unity at 1 MHz. Instead, for observation based limits, the ratio is around $10^{-24}$ still at 1 MHz.


\paragraph{Exploring the general DRs.}\mbox{} \\


Having caught a glimpse of what might happen, we now look at
a more general DR. When ${V}_{0}\neq 0$, for convenience and without loss of generality, we impose light propagating along the z axis ($k_1 = k_2 = 0$) that is  along the line of sight of the source. Incidentally, the group velocity has only a single component, and thus being unidimensional, there is no need to determine $|{\vec v_g}|$. We get from Eq. (\ref{CFJ DR}) 

\begin{gather}
\omega^{4}-\left(2k_{3}^{2}+V_{1}^{2}+V_{2}^{2}+V_{3}^{2}\right)\omega^{2}+2V_{0}V_{3}k_{3}\omega + \nonumber\\
\qquad k_{3}^{4}+\left(V_{1}^{2}+V_{2}^{2}-V_{0}^{2}\right)k_{3}^{2}=0~.
\label{omegaquattro}
\end{gather}

There are some interesting combinations of parameters to consider. The linear term impedes reduction to a quadratic equation. Hence, the components $V_{0}$ and $V_{3}$ will be inspected closely.


\paragraph{Non-zero time component of the breaking vector.}


We pose $V_{0}$, $V_1$ and $V_2$, different from zero, while $V_{3}=0$. In this case, we have\footnote{If we take $V_{0}=0$ in Eq. (\ref{ventisette}), the solution reads 
\beq
\bar{\omega}^{2}=\frac{2{\bar k}_3^2 + 1 \pm 1}{2}~. 
\label{footeq24}
\eeq
}

\begin{gather}
\bar{\omega}^{2}=\frac{2\bar{k}_{3}^{2}+1\pm\sqrt{1+4\bar{V}_{0}^{2}\bar{k}_{3}^{2}}}{2}~,
\label{ventisette}
\end{gather}
where we have rescaled the quantities as 

\beq
           \bar{\omega} =  \frac{\omega}{|\vec{V}|}~,
~~~~~~~~~~~~~~~~~\bar{V}_0  =  \frac{V_0}{|\vec{V}|}~,
~~~~~~~~~~~~~~~~~|\bar{k}_3| = \frac{k_3}{|\vec{V}|}~,\label{RenOmegaRenK}
\eeq

and where

\beq
{ |{\vec V}| = (V_1^2 + V_2^2)^{1/2}}~.
\label{RenV0}
\eeq

For the plus sign, the { right-hand side} of Eq. (\ref{ventisette}) is always positive, and thus we take the square root of this expression, derive and obtain the group velocity

\beq
v_{g+}=\frac{\bar{k_{3}}\left(1+{\displaystyle \frac{\bar{V}_{0}^2}{\sqrt{1+4\bar{V}_{0}^{2}\bar{k}_{3}^{2}}}}\right)}{{\displaystyle \sqrt{\bar{k}_{3}^{2}+\frac{1+\sqrt{1+4\bar{V}_{0}^{2}\bar{k}_{3}^{2}}}{2}}}}~,
\label{vg+}
\eeq

Under the same positive sign condition on Eq. (\ref{ventisette}), the group velocity $v_{g+}$ is never super-luminal, and frequencies are always real.

For the minus sign, the group velocity is 

\beq
v_{g-}=\frac{\bar{k_{3}}\left(1-{\displaystyle \frac{{ \bar{V}_{0}^2}}{\sqrt{1+4\bar{V}_{0}^{2}\bar{k}_{3}^{2}}}}\right)}{{\displaystyle \sqrt{\bar{k}_{3}^{2}+\frac{1-\sqrt{1+4\bar{V}_{0}^{2}\bar{k}_{3}^{2}}}{2}}}}~.
\label{vg-}
\eeq

Under the minus sign condition in Eq. (\ref{ventisette}), care is to be exerted. For a time-like breaking vector

\beq
V_{0}^{2}>|{\vec V}|^2 \Rightarrow\frac{V_{0}^{2}}{|{\vec V}|^2} = \bar{V}_0^2 >1~, 
\eeq

imaginary frequencies arise, from Eq. (\ref{ventisette}), if 

\beq
\bar{k}_{3}^{2} < \bar{V}_{0}^{2}  - 1~, 
\label{imfr1}
\eeq

and real frequencies occur, from Eq. (\ref{ventisette}), for 

\beq
\bar{k}_{3}^{2} \geq \bar{V}_{0}^{2}  - 1~.
\label{imfr1V0}
\eeq

When $\bar{k}_{3}$ is real, then $\bar{k}_{3}^2$ is positive; thus, for a space-like or light-like breaking vector, frequencies stay always real. 

Still for the minus sign in Eq. (\ref{ventisette}), we work out the group velocity in terms of $\omega$, keeping $V_3 = 0$. Using Eq. (\ref{omegaquattro}),
we write

\beq
2\omega_{\pm}^{2}=2k_{3}^{2}+|{\vec V}|^2\pm|{\vec V}|^2\sqrt{1+4\frac{V_{0}^{2}}{|{\vec V}|^{4}}k_{3}^{2}}~.
\label{casonuovo}
\eeq

However, $k_{3}$ is small if we are interested in the low frequency regime and 
${\displaystyle \frac{V_{0}^{2}}{|{\vec V}|^{2}}}\ll 1$ can
be assumed for a space-like $V^\mu$; thus

\beq
2\omega_{\pm}^{2}\sim2k_{3}^{2}+|{\vec V}|^2\pm|{\vec V}|^2\left(1+2\frac{V_{0}^{2}}{|{\vec V}|^{4}}k_{3}^{2}\right)
~,\label{Sol gen class 1}
\eeq

and so

\begin{eqnarray}
\omega_{\pm} & = & \left[k_{3}^{2}+\frac{|{\vec V}|^2}{2}\pm\left(\frac{|{\vec V}|^2}{2}+\frac{V_{0}^{2}}{|{\vec V}|^2}k_{3}^{2}\right)\right]^{\frac{1}{2}}=\nonumber \\
 & & \left[\left(1\pm\frac{V_{0}^{2}}{|{\vec V}|^2}\right)k_{3}^{2}+\frac{|{\vec V}|^2}{2}\pm\frac{|{\vec V}|^2}{2}\right]^{\frac{1}{2}}~.
\end{eqnarray}

Therefore, one root is 

\beq
\omega_{+}=\left({ \alpha}k_{3}^{2}+|{\vec V}|^2\right)^{\frac{1}{2}}~,\qquad \alpha=1+\frac{V_{0}^{2}}{|{\vec V}|^2},
\label{omega+34}
\eeq
where we have a dispersive behaviour with the parameter $|{\vec V}|$ acting once more as the mass of the photon, or else

\beq
\omega_{-}=\left (1-\frac{V_{0}^{2}}{|{\vec V}|^2}\right)^{1/2} |k_{3}|~,
\eeq
that is a dispersionless behaviour. When setting $V_{0}=0$, such that the parameter $\alpha$ reduces to
unity, we recover the Maxwell\-ian behaviour.

For the group velocities, from  Eq. (\ref{omega+34}), $k_{3}$ can
be explicitly written as

\beq
k_{3}=\frac{\omega_+}{\alpha^{1/2}}\left(1-\frac{|{\vec V}|^2}{\omega_+^{2}}\right)^{1/2},
\eeq
thus 
\begin{eqnarray}
v_{g+}=\frac{d\omega_+}{dk_3}=\frac{\alpha k_3}{\omega_+}&=&\alpha^{\frac{1}{2}}\left(1-\frac{|{\vec V}|^2}{\omega_+^2}\right)^{\frac{1}{2}}
= \nonumber \\
& & \alpha^{\frac{1}{2}}\left(1-\frac{|{\vec V}|^2}{2\omega_+^2}\right)+ \mathcal {O} \left(\frac{|{\vec V}|}{\omega}\right)^4~.
\label{Sol gen 1}
\end{eqnarray}

The other solution yields\footnote{Setting $V_{0}=0$, this result equals that of Eq. (\ref{drabalpha}) for ${ p} = -1$ and  
$\theta=\pi/2$ that is propagation along the $z$ axis.}

\beq
v_{g-}=1-\frac{V_{0}^{2}}{|{\vec V}|^2}~.
\label{Sol gen 2}
\eeq

We emphasise the domain of Eqs. (\ref{Sol gen 1},\ref{Sol gen 2}) cease when high frequencies and a time-like LSV vector are both considered.    
 
Here we obtain similar solutions to Eqs. (\ref{modvgcostheta0},\ref{modvgcostheta1}), differing by a factor depending on the time component of the CFJ breaking vector. However, this
coefficient is not trivial, and it offers some quite interesting features. 

The group velocity from Eq. { (\ref{Sol gen 2})} is never
super-luminal if $V_{\mu}$ is space-like. However, since $\alpha =1+{\displaystyle \frac{V_{0}^{2}}{|{\vec V}|^2}}$,
there is such a chance for the group velocity associated with Eq. { (\ref{Sol gen 1})}. It occurs for 

\beq
\sqrt{2}\omega_+>\frac{|{\vec V}|^2}{|V_0|}\left(1+{\displaystyle \frac{V_0^2}{|{\vec V}|^2}}\right)^{1/2}~.
\label{superlum}
\eeq

This is not surprising since it has been shown that $V_{0}$ might be associated to super-luminal modes. Setting $V_{0}=0$, we enforce luminal or sub-luminal speeds.  


 \paragraph{Presence of all breaking vector components and $V^{\mu}$ light-like.} 


When all parameters differ from zero in Eq. (\ref{omegaquattro}), it is obviously the most complex case. Nevertheless, we can comment specific solutions. 

We suppose the vector $V^{\mu}$ being light-like.
%
%
%
%
%
%

Thereby, we have $V^{2}=0\Rightarrow\left(V^{0}\right)^{2}=|\vec{V}|^{2}\Rightarrow |V^{0}|=|\vec{V}|\Rightarrow V^{0}=\pm|\vec{V}|$ (we choose $V^{0}=|\vec{V}|$, without loss of generality). The DR from Eq. (\ref{CFJ DR}) and from Eq. (\ref{DR-class2}) for $H, M \rightarrow 0$ reads 

\begin{equation}
k^{4}+V^{2}k^{2}-\left(V\cdot k\right)^{2}= \left(k^{2}\right)^{2}- \left(V\cdot k\right)^{2}=0 \Rightarrow|k^{2}|=|V\cdot k|~.
\end{equation}

When considering $k^{2}\geq0$, thus $|k^{2}|=k^{2}$, part of the tachyonic modes are excluded, but others survive, as shown below.  
We have

\begin{equation}
k^{2}=\omega^{2}-|\vec{k}|^{2}=|V\cdot k|=|V^{0}\omega-\vec{V}\cdot\vec{k}|~.
\end{equation}

Hence, two cases arise, for the positiveness of $k^{2}\geq0$:
\begin{itemize}
\item {Case 1: $V^{0}\omega-\vec{V}\cdot\vec{k}\geq0\Rightarrow\omega^{2}-|\vec{k}|^{2}=V^{0}\omega-\vec{V}\cdot\vec{k}$~,}
\item {Case 2: $V^{0}\omega-\vec{V}\cdot\vec{k}\leq0\Rightarrow\omega^{2}-|\vec{k}|^{2}=-V^{0}\omega+\vec{V}\cdot\vec{k}$~.}
\end{itemize}

For case 1, we have

\begin{eqnarray}
\omega^{2} - V^{0}\omega & & -\left(\vec{k}^{2}-\vec{V}\cdot\vec{k}\right) =  \nonumber \\ 
\omega^{2} - V^{0}\omega & & - |\vec{k}|\left(|\vec{k}|-|\vec{V}|\cos\theta\right) = 0 ~,
\label{caso1}
\end{eqnarray}

the solutions of which are
\beq
\omega_{1} = \frac{V^{0}}{2}\pm\sqrt{\left(\frac{V^{0}}{2}\right)^{2}+ |\vec{k}|\left(|\vec{k}|-|\vec{V}|\cos\theta\right)}~, 
\label{omegapm1}
\eeq

and since $V^{0}=|\vec{V}|$, we finally get

\beq
\omega_{1} = \frac{|\vec{V}|}{2}\pm\sqrt{\left(\frac{|\vec{V}|}{2}\right)^{2}+ |\vec{k}|\left(|\vec{k}|-|\vec{V}|\cos\theta\right)}~. 
\label{omegapm1fin}
\eeq

We consider only a positive radicand and exclude negative frequencies.  

Similarly for case 2, we have the following equation and solutions

\begin{eqnarray}
\omega^{2} + V^{0}\omega & & -\left(\vec{k}^{2}+\vec{V}\cdot\vec{k}\right) =  \nonumber \\ 
\omega^{2} + V^{0}\omega & & - |\vec{k}|\left(|\vec{k}|+|\vec{V}|\cos\theta\right) = 0 ~,
\label{caso2}
\end{eqnarray}

\beq
\omega_{2} = - \frac{|\vec{V}|}{2} + \sqrt{\left(\frac{|\vec{V}|}{2}\right)^{2}+ |\vec{k}|\left(|\vec{k}|+|\vec{V}|\cos\theta\right)}~, 
\label{omegapm2fin}
\eeq

having excluded negative frequencies.
  
We now discuss the current bounds on the value of the breaking vector in Sect. \ref{upperlim} in SI units.  
 In the yet unexplored low radio frequency spectrum \cite{bebosp2017}, a frequency of $10^5$ Hz 
and a wavelength $\lambda$ of $3\times 10^{3}$ m results in $|\vec{k}| \hbar c={\displaystyle \frac{2\pi}{\lambda}} \hbar c \sim 6.3 \times 10^{-30}$ J, while in the gamma-ray regime, a wavelength
$\lambda$ of $3\times 10^{-11}$ m results in $|\vec{k}| \hbar c={\displaystyle \frac{2\pi}{\lambda}} \hbar c \sim 6.3 10^{-16}$ J. Spanning the domains of the parameters $\vec V$ and ${\vec k}$, we cannot assure the positiviness of the factor $|\vec{k}|-|\vec{V}|\cos\theta$ in Eq. 
(\ref{omegapm1fin}). Moving toward smaller but somewhat less reliable astrophysical upper limits, we insure such positiveness. The non-negligible price to pay is that the photon effective mass and the perturbation vector decrease and their measurements could be confronted with the Heisenberg limit, see Sect. \ref{upperlim}. This holds especially for low frequencies around and below $10^5$ Hz.

For case 1, for a positive radicand, we have 

\begin{equation}
\sqrt{\left(\frac{\vec{V}}{2}\right)^{2}+\vec{k}^{2}-\vec{V}\cdot\vec{k}}=\sqrt{\left(\frac{\vec{V}}{2}-\vec{k}\right)^{2}}=\left|\frac{\vec{V}}{2}-\vec{k}\right|~,\label{simplification}
\end{equation}

and the allowed solutions for $\omega_1$ are

\begin{equation}
\omega_{1a}= \frac{|\vec{V}|}{2} - \left|\frac{\vec{V}}{2}-\vec{k}\right|~;~~~~~~~~~~~~~~~~~\omega_{1b}=\frac{|\vec{V}|}{2} + \left|\frac{\vec{V}}{2}-\vec{k}\right|~. 
\label{omega1positive}
\end{equation}

For case 2 the allowed solutions for $\omega_2$ is only 

\begin{equation}
\omega_{2}= - \frac{|\vec{V}|}{2} + \left|\frac{\vec{V}}{2}+\vec{k}\right|~. 
\label{omega2positive}
\end{equation}

For Case 1 group velocity, by Eq. (\ref{caso1}) we get

\beq
2\omega d\omega - V^0 d\omega - 2k_i dk_i + V_i dk_i = 0
\eeq 

and thereby 

\beq
v_{gi} = \frac{d\omega}{dk_i} = \frac{k_i - {\ds \frac{V_i}{2}}}{\omega - {\ds \frac{V^0}{2}} }~;
\eeq

\beq
{\vec v}_{g} = \frac{{\vec k} - {\ds \frac{{\vec V}}{2}}}{\omega - {\ds \frac{V^0}{2}} }
= \frac{{\vec k} - {\ds \frac{{\vec V}}{2}}}{\omega - {\ds \frac{|\vec{V}|}{2}} }~.
\label{vggenerallightlike1}
\eeq

From the expressions of $\omega_{1a,b}$ we write 

\beq
\omega_{1a,b} - \frac{|\vec{V}|}{2} = \pm \left|\frac{\vec{V}}{2}-\vec{k}\right|~, 
\eeq

and evince that the absolute value of the group velocity is equal to unity.  
For Case 2 group velocity, by Eq. (\ref{caso2}) we get 

\beq
2\omega d\omega + V^0 d\omega - 2k_i dk_i - V_i dk_i = 0
\eeq 

and thereby 

\beq
v_{gi} = \frac{d\omega}{dk_i} = \frac{k_i + {\ds \frac{V_i}{2}}}{\omega + {\ds \frac{V^0}{2}} }~;
\eeq

\beq
{\vec v}_{g} = \frac{{\vec k} + {\ds \frac{{\vec V}}{2}}}{\omega + {\ds \frac{V^0}{2}} }
= \frac{{\vec k} + {\ds \frac{{\vec V}}{2}}}{\omega + {\ds \frac{|\vec{V}|}{2}} }~.
\label{vggenerallightlike2}
\eeq

From the expressions of $\omega_2$ we write 

\beq
\omega_2 + \frac{|\vec{V}|}{2} = \left|\frac{\vec{V}}{2}+\vec{k}\right|~, 
\eeq
  
and evince once more that the absolute value of the group velocity is equal to unity.  We thereby conclude that even when the frequency differs from $|\vec{k}|$, the group velocity is Maxwellian, for a light-like $V^\mu$.  

The most general case represented by Eq. (\ref{omegaquattro}) should be possibly dealt with a numerical treatment.  


\paragraph{Time delays.\label{timedelays}} \mbox{} \\


For better displaying the physical consequences of these results, we compute the time delay between two waves of different frequencies  \cite{db40}. In SI units, for a source at distance $l$ (Eq. (16) in \cite{bodshnsp2017})

\beq
\Delta t_{CFJ}= \frac {l  |\vec{V}|^2}{2c\hbar^2} \left(\frac{1}{\omega_{1}^{2}}-\frac{1}{\omega_{2}^{2}}\right) x~.
\label{time delay class 1}
\eeq
where $\hbar$ is the reduced Planck constant (also Dirac constant) and $x$ takes the value $1$ for 
Eq. (\ref{modvgcostheta0}), $1/4$ for Eq. (\ref{modvgcostheta1}) or $\alpha^{1/2}$ for Eq. (\ref{Sol gen 1}). Obviously, other values of $x$ are possible, when considering more general cases.

As time delays are inversely proportional to the square of the frequency, we perceive the existence of a massive photon, in presence of gauge invariance, emerging from the CFJ theory. Its mass value is proportional to the breaking parameter $|{\vec V}|$.  The comparison of 
Eq. (\ref{time delay class 1}) with the corresponding expression for the de Broglie-Proca (dBP) photon \cite{db40} 

\beq
\Delta t_{DBP}=\frac{l\ m_{\gamma}^{2}c^3}{2h^2}\left(\frac{1}{\omega_{1}^{2}}-\frac{1}{\omega_{2}^{2}}\right)~,\label{delay DBP}
\eeq

leads to the identity (Eq. (18) in \cite{bodshnsp2017})

\beq
m_{\gamma}= \frac{|{\vec V}|}{c^2} x~.
\eeq

We recall that Class 2 is just a rescaling of Class 1, where the correcting factor $1/(1-H-M)^2$
is extremely close to unity.

Finally, given the prominence of the delays of massive photon dispersion, either of dBP or CFJ type, at low frequencies, a swarm of nano-satellites operating in the sub-MHz region \cite{bebosp2017} appears a promising avenue for improving upper limits through the analysis of plasma dispersion.


\subsubsection{A quasi-de Broglie-Proca-like massive term. \label{sec:An-approach}} \mbox{} \\


A quasi-dBP-like term from the CPT-odd Lagrangian has been extracted \cite{bodshnsp2017}, but without giving details. Indeed, the interaction of the photon with the background gives rise to an effective mass for the photon, depending on the breaking vector $V^{\mu}$. As we will show,
this can be linked to the results we obtained from the DR applied to polarised fields.

We cast the CPT-odd Lagrangian, Eq. (\ref{CFJ_lagrangian}) in terms of the potentials

\begin{eqnarray}
L & = & \frac{1}{2}\left(\vec{\nabla}\phi+\dot{\vec{A}}\right)^{2}-\frac{1}{2}\left(\vec{\nabla}\times\vec{A}\right)^{2}+V_{0}\vec{A}\cdot\left(\vec{\nabla}\times\vec{A}\right)\nonumber \\
 &  & \qquad-\phi\vec{v}\cdot\left(\vec{\nabla}\times\vec{A}\right)-\vec{v}\cdot\left(\vec{A}\times\dot{\vec{A}}\right)-\left(\vec{V}\times\vec{A}\right)\cdot\vec{\nabla}\phi = \nonumber \\ 
& & \frac{1}{2}\left(\vec{\nabla}\phi+\dot{\vec{A}}\right)^{2}-\frac{1}{2}\left(\vec{\nabla}\times\vec{A}\right)^{2}+V_{0}\vec{A}\cdot\left(\vec{\nabla}\times\vec{A}\right)- \nonumber\\
& & ~~~~~~~2\vec{\nabla}\phi\cdot \left(\vec{V}\times\vec{A}\right)-\vec{V}\cdot\left(\vec{A}\times\dot{\vec{A}}\right)~.
\label{LCFJfields}
\end{eqnarray}

The scalar potential $\phi$ always appears through its gradient, implying that $\vec{\nabla}\phi$ is the true
degree of freedom. Further, in absence of time derivatives of this field, there isn't dynamics. In other words,
$\phi$  plays the role of an auxiliary field, which can be eliminated from the Lagrangian. We call 

\beq
\vec{\nabla}\phi=\vec{S}~,
\eeq
and rewrite the CPT-odd Lagrangian as

\begin{eqnarray}
L & = & \frac{1}{2}\left(\vec{S}+\dot{\vec{A}}-2\vec{V}\times\vec{A}\right)^{2}-2\left(\vec{V}\times\vec{A}\right)^{2}+2\dot{\vec{A}}\cdot\left(\vec{V}\cdot\vec{A}\right)- \nonumber \\
 &  & \qquad \frac{1}{2}\left(\vec{\nabla}\times\vec{A}\right)^{2}+V_{0}\vec{A}\cdot\left(\vec{\nabla}\times\vec{A}\right)-\vec{V}\cdot\left(\vec{A}\times\dot{\vec{A}}\right)~.
\end{eqnarray}

Defining $\chi$ as 
\beq
\chi=\vec{S}+\dot{\vec{A}}-2\vec{V}\times\vec{A}~,
\eeq
we get 
\begin{gather}
L=\frac{1}{2}\chi^{2}-2\left(\vec{V}\times\vec{A}\right)^{2}+\vec{V}\cdot\left(\vec{A}\times\dot{\vec{A}}\right)-\frac{1}{2}\left(\vec{\nabla}\times\vec{A}\right)^{2}+\nonumber\\
\qquad\qquad V_{0}\vec{A}\cdot\left(\vec{\nabla}\times\vec{A}\right)~.
\end{gather}

Passing through the Euler-Lagrange equations, we derive $\chi=0$. Therefore $\chi$ is cancelled out, and we are left with

\beq
L=\vec{V}\cdot\left(\vec{A}\times\dot{\vec{A}}\right)-2\left(\vec{V}\times\vec{A}\right)^{2}-\frac{1}{2}\left(\vec{\nabla}\times\vec{A}\right)^{2}+V_{0}\vec{A}\cdot\left(\vec{\nabla}\times\vec{A}\right).
\eeq

Since the vector potential $\vec{A}$ does not appear with derivatives, further elaboration leads to

\begin{eqnarray}
L &= &\vec{V}\cdot\left(\vec{A}\times\dot{\vec{A}}\right)-2M_{kn}\left(\vec{V}\right)A_{k}A_{n}-\frac{1}{2}\left(\vec{\nabla}\times\vec{A}\right)^{2}+ \nonumber \\
& & V_{0}\vec{A}\cdot\left(\vec{\nabla}\times\vec{A}\right)~,
\end{eqnarray}
where
\beq
M_{kn}\left(\vec{V}\right)= |\vec{V}|^{2}\delta_{kn}-V_{k}V_{n}~.
\eeq
which is a symmetric matrix, thereby diagonalisable

\beq
-2M_{kn}\left(\vec{V}\right)A_{k}A_{n} = - 2A^{T}MA = -2A^{T}R^{T}RMR^{T}RA~,
\eeq
where $R\in SO\left(3\right)$ diagonalises $M$ and $A^{T}$ being the latter the transposed potential vector. We label

\beq
\tilde{M}=RMR^{T}=\left(\begin{array}{ccc}
m_{1} & 0 & 0\\
0 & m_{2} & 0\\
0 & 0 & m_{3}
\end{array}\right)~.
\eeq

and get 
\begin{gather}
\mbox{det}~\tilde{M}=0\Rightarrow m_{1}=0~,\\
\mbox{Tr}~\tilde{M}=m_{1}+m_{2}+m_{3}\Rightarrow { m_{2}=m_{3}}=|\vec{V}|^{2}~.
\end{gather}

Therefore the term
\beq
\tilde{A}_{i}\tilde{M}_{ij}\tilde{A}_{j}= |\vec{V}|^{2}\tilde{A}_{2}^{2}+|\vec{V}|^{2}\tilde{A}_{3}^{2}~,\label{massive term}
\eeq
is a dBP term as we wanted (Eq. 21 in \cite{bodshnsp2017}. The role of the mass is played by the modulus of the vector $\vec{V}$. A remarkable difference lies in the gauge independency of the CFJ massive term.


\subsection{The CPT-even sector and the $k_{F}$ tensor: classes 3 and 4}


For the CPT-even sector, in \cite{bebegahnle2015} the authors investigate the $k_{F}$-term from 
SME, focusing on how the Fermionic condensates affect the physics of photons and photinos.

In the $k_{F}$ tensor model Lagrangian, the LSV term is 

\beq
L_3=\left(k_{F}\right)_{\mu\nu\alpha\beta}F^{\mu\nu}F^{\alpha\beta}, \label{L3}
\eeq
\label{sixtyseven}
where $\left(k_{F}\right)_{\mu\nu\alpha\beta}$ is double traceless. The $k_F$ tensor, see \ref{appeven}, is written in terms of a single Bosonic vector $\xi_{\mu}$ which signals LSV

\beq
\left(k_{F}\right)_{\mu\nu\alpha\beta}=\frac{1}{2}\left(\eta_{\mu\alpha}{\kappa}_{\nu\beta}-\eta_{\mu\beta}{\kappa}_{\nu\alpha}+\eta_{\nu\beta}{\kappa}_{\mu\alpha}-\eta_{\nu\alpha}{\kappa}_{\mu\beta}\right)~, 
\eeq
being 

\beq
{\kappa}_{\alpha\beta}=\xi_{\alpha}\xi_{\beta}-\eta_{\alpha\beta}\frac{\xi_{\rho}\xi^{\rho}}{4}~. 
\eeq

As it is mentioned in \ref{appeven}, in Eqs. (\ref{a1},\ref{kF}), we choose $k_F$ to be given according to the non-birefringent {\it Ansatz}, as discussed in \cite{bekakl2009,baileykostelecky2004}.


\subsubsection{Class 3: $k_F$ model}


Following \cite{bebegahnle2015,bebegahn2013}, the DR for the photon reads (Eq. (8) in \cite{bodshnsp2017})

\beq
\omega^{2}-\left(1+\rho+\sigma\right)^{2}|{\vec k}|^{2}=0~,
\label{DRclass3}
\eeq
where

\begin{eqnarray}
\rho & = & \frac{1}{2}\bar{K}_{\alpha}^{\alpha}\label{rho}~,\\
 \sigma & = & \frac{1}{2}\bar{K}_{\alpha\beta}\bar{K}^{\alpha\beta}-\rho^{2}\label{sigma}~,\\
\bar{K}^{\alpha\beta} & = & t^{\alpha\beta}t^{\mu\nu}\frac{k_{\mu}k_{\nu}}{|{\vec k}|^2}.\label{Ktilde}~,
\end{eqnarray}

being $t^{\mu\nu}$ a constant symmetric tensor corresponding to the condensation of the background scalar present in the background super-multiplet that describes $k_F$-LoSy breaking. This tensor is related to the $k_F$ term of Eq. (\ref{sixtyseven}) in a SuSy scenario; its origin is explained in \cite{bebegahnle2015}. It is worthwhile recalling that for such a tensor, the absence of the time component excludes the appearance of tachyons and ghosts. Therefore, in Eq. (\ref{Ktilde})
we take only the $ij$ components

\beq
\bar{K}^{ij}=t^{ij}t^{mn}\frac{k_{m}k_{n}}{|{\vec k}|^2}~,\label{Ktildespatial}
\eeq

Moreover, the tensor $t$ is always symmetric, hence we can always diagonalise it. 

The simplest case occurs when the breaking tensor is a multiple of the identity. Then, Eq. (\ref{Ktildespatial}) becomes
\beq
\bar{K}^{ij}=t^{2}\delta^{ij}\delta^{mn}\frac{k_{m}k_{n}}{|{\vec k}|^2}=t^{2}\delta^{ij}~.
\eeq

This means that both $\rho$ and $\sigma$ are independent of $\vec{k}$ or $\omega$ and that the factor in front of ${\vec k}^{2}$
in Eq. (\ref{DRclass3}) carries no functional dependence. Therefore, we have a situation where the vacuum acts like a medium,
whose refraction index is given by
\beq
n=\left(1+\rho+\sigma\right)^{-1}~.
\eeq

The most general case occurs when $t^{ij}$ is diagonal and not traceless. Then, we have 

\beq
t^{ij}=t_{i}\delta^{ij}~,
\label{tij}
\eeq
where we have left aside the Einstein summation rule. Equation (\ref{Ktildespatial}) is rewritten as

\beq
\bar{K}^{ij}=t_{i}\delta^{ij}\left(t_{m}\delta^{mn}\frac{k_{m}k_{n}}{|{\vec k}|^2}\right)~,
\eeq
where the term within the round brackets is
 
\begin{gather}
\frac{1}{|{\vec k}|^2}tr \left[\left(\begin{array}{ccc}
t_{1} & 0 & 0\\
0 & t_{2} & 0\\
0 & 0 & t_{3}
\end{array}\right)\left(\begin{array}{ccc}
k_{1}^{2} & k_{1}k_{2} & k_{1}k_{3}\\
k_{2}k_{1} & k_{2}^{2} & k_{2}k_{3}\\
k_{3}k_{1} & k_{3}k_{2} & k_{3}^{2}
\end{array}\right)\right]\nonumber \\ 
\qquad\qquad\qquad=\frac{t_{1}k_{1}^{2}}{|{\vec k}|^2}+\frac{t_{2}k_{2}^{2}}{|{\vec k}|^2}+\frac{t_{3}k_{3}^{2}}{|{\vec k}|^2}:=P { ({\vec k})} = t_i\frac
{k_i^2}{|{\vec k|}^2}~.
\label{matrice}
\end{gather}

Now, using Eq. (\ref{rho}), Eq. (\ref{sigma}) is transformed into
 
\begin{eqnarray}
\sigma & = & \frac{1}{2}\mbox{tr}\left(\bar{K}^{2}\right)-\left(\frac{1}{2}\mbox{tr}\bar{K}\right)^{2}~.\label{sigma square}
\end{eqnarray}

Since
\beq
\frac{1}{2}\mbox{tr}\bar{K}=\frac{1}{2}P{ ({\vec k})}\left(t_{1}+t_{2}+t_{3}\right)~,
\eeq

and 

\beq
\text{tr}\left(\bar{K}^{2}\right)
=P^{2}\left(t_{1}^{2}+t_{2}^{2}+t_{3}^{2}\right):=P^{2}{ ({\vec k})}F^{2}~,
\eeq

Eq. (\ref{sigma square}) becomes

\beq
\sigma  = \left[\frac{F^{2}}{2}-\frac{\left(t_{1}+t_{2}+t_{3}\right)^2}{4}\right]P^{2}{ ({\vec k})}~.
\eeq

Discarding the negative frequency solution, from Eq. (\ref{DRclass3}), we are left with 

\beq
\omega=\left(1+\rho+\sigma\right)|{\vec k}|~,
\eeq
which explicitly becomes

\begin{eqnarray}
\omega & = & \left\{ 1+\left[\frac{1}{2}\left(t_{1}+t_{2}+t_{3}\right)+\frac{F^{2}}{2}-\frac{\left(t_{1}+t_{2}+t_{3}\right)^2}{4}\right]
P\right\} |{\vec k}|\nonumber \\
 & := & \left(1+C\  P\right)|{\vec k}|~,
\label{massless3}
\end{eqnarray}

where $C$ depends exclusively on the $t_i$ parameters. The dependency on $\vec k$ goes through $P$, Eq. (\ref{matrice}). Considering the anisotropy represented by the eigenvalues $t_{i}$ of Eq. (\ref{tij}), we compute the group velocity along the i$^{th}$ space direction

\beq 
v_{gi} = \frac{\partial \omega}{\partial k_i}~,
\eeq

and thereby, we find

\beq 
v_{gi} = \frac{k_i}{|{\vec k}|} \left( 1 + 2 C t_i +  C t_j\frac{k_i^2}{|{\vec k}|}\right )~, 
\eeq
where summation does not run over $i$ ($i$ fixed),  but over $j$. 

Finally, we get the group velocity 

\beq
|{\vec v}_{g}|^2= 1 + 6 C t_i \frac{k_i^2}{|{\vec k}|^2}+ {\mathcal O}(t^2)~,
\label{vg3}
\eeq
where summation runs over the index $i$.

This shows a non-Maxwellian behaviour, $|{\vec v}_{g}| \neq 1 $, whenever the second left-hand side term differs from zero. We observe that there is a frequency dependency, but absence of mass since, Eq. (\ref{massless3}), $|{\vec k}| = 0$ implies $\omega = 0$. 
The frequency never becomes complex, while super-luminal velocities may appear if $C t_i k_i^2$ in Eq. (\ref{vg3}) is positive. 
   The parameters $t_{}$ are suppressed by powers of the Planck energy, so they are very small. This justifies the truncation in Eq. (\ref{vg3}). The value depends on the constraints of such parameters.


\subsubsection{Class 4: $k_F$ model and SuSy breaking}


As we did for Class 2, we proceed towards an effective photonic Lagrangian for Class 4, by integrating out the photino sector.
The resulting Lagrangian reads \cite{bebegahnle2015}

\beq
L_{4}=-\frac{1}{4}F_{\mu\nu}F^{\mu\nu}+\frac{{ r}}{2}\chi_{\mu\nu}F_{\kappa}^{\mu}F^{\nu\kappa}+\frac{{ s}}{2}\chi_{\mu\nu}\partial_{\alpha}F^{\alpha\mu}\partial_{\beta}F^{\beta\nu}~,
\label{Lclass4}
\eeq

The $\chi^{\alpha\beta}$ tensor is linearly related to the breaking tensor $k_F$, as it has been shown in the Appendix B of
 \cite{bebegahnle2015}. Also, according to the results of Sects. 2, 4 of the same reference, the - mass$^{-2}$ - parameter s corresponds 
to the (scalar) condensate of the Fermions present in the background SuSy multiplet responsible for the LoSy violation, 
where $r$ is a dimensionless coefficient, estimated as $r = -32$ \cite{bebegahnle2015}. 
The term with coefficient $s$ in $L_4$ corresponds to a dimension-6 operator and, in a context without SuSy, it appears in the photon sector of the non-minimal SME \cite{kome09,kosteleckyrussell2011}. More recently 
\cite{casana-ferreira-lisboasantos-dossantos-schreck-2018}, an analysis of causality and propagation properties stemming from the dimension-6 term above was carried out.

The DR reads (Eq. 10 in \cite{bodshnsp2017}, see \ref{appeven}

\begin{gather}
s
\chi k^4 - (1 - r \chi + s \chi^{\alpha\beta}k_{\alpha}k_{\beta})k^2+ 3 r \chi^{\alpha\beta}k_{\alpha}k_{\beta}=0~,
\label{Class4-DR-almost}
\end{gather}
where $\chi=\chi_{\mu}^{\mu}$.

Similarly to Class 2, the tensor $\chi_{\alpha\beta}$ is symmetric and thus diagonalisable. If the temporal components linked to  super-luminal and ghost solutions are suppressed ($\chi^{00}= \chi^{0i}=0$), we get 
 
\beq
\chi =  \chi^1_{~1}+\chi^2_{~2}+\chi^3_{~3} : =  \chi_{1}+\chi_{2}+\chi_{3}
\eeq

where again, we disregard Einstein summation rule for the $i$ index. For 

\beq
\chi^{\alpha\beta}k_{\alpha}k_{\beta}= { -}\chi_{1}k_{1}^{2}{ -}\chi_{2}k_{2}^{2}{ -}\chi_{3}k_{3}^{2}:=D{(k)}~, 
\eeq

we get
\begin{gather}
s \chi\left(\omega^{2}-|{\vec k}|^2\right)^{2} - \left( 1 - r\chi + sD \right)\left(\omega^{2}-|{\vec k}|^2\right)
+  3 r D = 0~.
\label{Class4-DR}
\end{gather}

Expanding for $\omega$ 

\begin{gather}
s\chi\omega^{4} -  \left(1 - r\chi + 2 s\chi|{\vec k}|^2 + sD \right)\omega^{2} + \nonumber \\
s\chi|{\vec k}|^4  + \left(1 - r\chi + sD\right)|{\vec k}|^2 + 3 r D=0~.
\label{mass4}
\end{gather}

Rather than solving the fourth order equation, we derive the group velocity at first order in $\chi_i$
 
\beq
v_{gi} = \frac{k_i}{\omega} - (3 { r} + { s}   |{\vec k}|^2 )\chi_i \frac{k_i}{\omega} + { s}\omega\chi_i k_i + {\mathcal O}(\chi^2)~, 
\label{rather}
\eeq

where there isn't summation over the index $i$. Finally, we get 

\beq
|{\vec v}_{g}|^2 =  \frac{|{\vec k}|^2}{\omega^2} + \frac{2(3 r +  s   |{\vec k}|^2 )D}{\omega^2} - 2sD 
+ {\mathcal O}(\chi^2)~, 
\label{vg4massiva}
\eeq

The behaviour with frequency of the group velocity is also proportional to the inverse of the frequency squared, as for the dBP massive photon.

Conversely to Class 3, here the integration of the photino leads to a massive photon, evinced from $\omega \neq 0$ for $k=0$, Eq. (\ref{mass4}). This was undetected in our previous work \cite{bodshnsp2017}. The photon mass comes out as 
 
\beq
m_\gamma = \left ( \frac{1-r\chi}{s\chi}\right)^{1/2}~.
\label{massclass4}
\eeq

In \cite{kome09,casana-ferreira-lisboasantos-dossantos-schreck-2018}, there isn't any estimate on the s-parameter. 
In \cite{kosteleckyrussell2011}, besides assessing the dimensionless $k_F$ as $10^{-18}$, the authors present Table XV of the estimates on the parameters associated to dimension-6 operators. They are based on observations of astrophysical dispersion and bi-refringence. Considering our DR of Eq. (\ref{Class4-DR-almost}), the PDG \cite{tanabashietal2018} photon mass limit of 
$5.6\times 10^{-19}$ eV/c$^2$ and the estimate in Appendix B of \cite{bebegahnle2015}, for $\chi = \sqrt {k_F} 10^{-9})$, $\sqrt{1/s}$ is evaluated as $1.8 \times 10^{-24}$ eV/c$^2$.

Super-luminal velocities may be generated and  
$\omega^2$ becomes complex if, referring to Eq. (\ref{rather})

\beq
\left(1 - r\chi + sD \right )^2 - 12 rs\chi D<0~.
\eeq


\section{Bi-refringence in CPT-odd classes\label{tririf}}


For CPT-odd classes, the determination of the DRs in terms of the fields provides a fruitful outcome, since it relates the solutions to the physical polarisations of the fields themselves. This approach must obviously reproduce compatible results with those obtained with the potentials. However, the physical interpretation of said results should be clearer in this new approach.

We consider the wave propagating along one space component of the breaking vector $V^{\mu}$. Without loss of generality, we pose  
$\vec{V}=V\hat{\vec z}$ and $\vec{k}=k \hat{\vec z}$. The fields ($\vec{e}$, $\vec{b}$ of the photon) are then written as
 
\begin{eqnarray}
{ \vec{e}} & = & { \vec{e}}_{0}e^{i\left(kz-\omega t\right)}~, \\
{ {\vec b}} & = & { {\vec b}}_{0}e^{i\left(kz-\omega t\right)}~,
\end{eqnarray}
where ${ \vec{e}}_{0}$ and ${ {\vec b}}_{0}$ are complex vectors  

\begin{eqnarray}
{ \vec{e}}_{0} & = & { \vec{e}}_{0R}+i{ \vec{e}}_{0I}~,\\
{ {\vec b}}_{0} & = & { {\vec b}}_{0R}+i{ {\vec b}}_{0I}~,
\end{eqnarray}
the subscripts $R$ and $I$ standing for the real and imaginary parts, respectively. 
The actual fields are the real parts of ${ \vec{e}}$ and ${\vec b}$

\begin{eqnarray}
{ \vec{e}} & = & { \vec{e}}_{0R}\cos\left(kz-\omega t\right)-{ \vec{e}}_{0I}\sin\left(kz-\omega t\right)~,\\
{ {\vec b}} & = & { {\vec b}}_{0R}\cos\left(kz-\omega t\right)-{ {\vec b}}_{0I}\sin\left(kz-\omega t\right)~.
\end{eqnarray}

From the field equations \cite{cafija90}, the following relations emerge

\begin{gather}
\vec{k}\cdot{ \vec{e}}_{0R}+\vec{V}\cdot{ {\vec b}}_{0I}=0~,\label{prima della lista}\\
\vec{k}\cdot{ \vec{e}}_{0I}-\vec{V}\cdot{ {\vec b}}_{0R}=0~,\\
\vec{k}\times{ \vec{e}}_{0R}=\omega{ {\vec b}}_{0R}~,\label{rel 1}\\
\vec{k}\times{ \vec{e}}_{0I}=\omega{ {\vec b}}_{0I}~,\label{rel 2}\\
\vec{k}\cdot{ {\vec b}}_{0R}=\vec{k}\cdot{ {\vec b}}_{0I}=0~,\\
-\vec{k}\times{ {\vec b}}_{0R}-V_{0}{ {\vec b}}_{0I}+\vec{V}\times{ \vec{e}}_{0I}=\omega{ \vec{e}}_{0R}\label{rel 3}~,\\
-\vec{k}\times{ {\vec b}}_{0I}+V_{0}{ {\vec b}}_{0R}-\vec{V}\times{ \vec{e}}_{0R}=\omega{ \vec{e}}_{0I}.\label{rel 4}~.
\end{gather}

From the above relations, recalling that both $\vec{k}$ and $\vec{V}$
are along the $\hat{\vec z }$ axis, we obtain that ${ \vec{e}}_{0R}$ and
${ \vec{e}}_{0I}$ are transverse. They develop longitudinal components
only if $\vec{V}\cdot{ {\vec b}}_{0R}$ and $\vec{V}\cdot{ {\vec b}}_{0I}$
are non vanishing.

Dealing with a transverse ${ \vec{e}}_{0}$, we consider a circularly
polarised wave

\begin{eqnarray}
{ \vec{e}}_{0R} & = & { e}_{0}\hat{{\vec x}}\label{circ +}~,\\
{ \vec{e}}_{0I} & = & \xi { e}_{0}\hat{{\vec y}}~,\label{circ -}
\end{eqnarray}
implying

\beq
{ \vec{e}}_{0}={ e}_{0}{ \left(\hat{{\vec x}}+i\xi\hat{{\vec y}}\right)}~,
\eeq
with $\xi=\pm1$ indicating right- ($+1$) or left-handed ($-1$) polarisation. 
Using
Eqs. (\ref{rel 1}, \ref{rel 2}, \ref{rel 3}-\ref{circ -}), 
the following dispersion is written

\beq
\omega^{2}+\xi V\omega-k^{2}-\xi V_{0}k=0~,\label{group velocity tri-refringence}
\eeq
from which a polarisation dependent group velocity can be attained
\beq
v_{g}=\frac{1}{2\omega+\xi |{\vec V}|}\sqrt{\left(2\omega+\xi |{\vec V}|\right)^{2}+V_{0}^{2}-|{\vec V}|^2}.
\label{vgpolcirc}
\eeq

Up to Eq, (\ref{group velocity tri-refringence}), we have not specified the space-time character of the background vector $V_\mu$. However, Eq. 
(\ref{vgpolcirc}) shows $v_g > 1$, if $V_\mu$ is time-like. So, to avoid super-luminal effects, we restrict $V_\mu$ to be a space- or light-like four-vector. In the former case $v_g < 1$, in the latter $v_g =1$.

The group velocity dependency on the two value-handed\-ness is known as bi-refringence. Incidentally, the
group velocity from Eq. (\ref{vgpolcirc}) can be expressed
in terms of the wave number $k$

\beq
v_{g}\left(k\right)=\left(2k+\xi V_{0}\right)\left[\left(2k+\xi V_{0}\right)^{2}-V_{0}^{2}+V^{2}\right]~.\label{eq59}
\eeq

For a situation of linear polarisation (${\vec k}$ and ${\vec V}$ being parallel), if we consider $V^{\mu}$ light-like, we have

\begin{eqnarray}
{ \vec{e}}_{0R} & = & { e}_{0}\hat{{\vec x}}~,\\
{ \vec{e}}_{0I} & = & 0~.
\end{eqnarray}

In this case, Eqs. (\ref{prima della lista}-\ref{rel 4}) lead to 

\begin{eqnarray}
{ {\vec b}}_{0I} & = & 0~,\\
{ {\vec b}}_{0R} & = & \frac{k}{\omega}{ e}_{0}\hat{{\vec y}}~,
\end{eqnarray}

and the group velocity turns out to be 

\beq
v_{g}=1~,
\eeq
showing that to the linear polarisation is associated a different $v_{g}$. 

One might be persuaded{, as we initially were,} that this result entails the property of tri-refringence,
because with the same wave vector as in the case of circular polarization, we
get a different $v_g$, namely, $v_g = 1$. And tri-refringence actually means 
three distinct refraction indices for the same wave vector. However, the linear
polarisation and the result $v_g = 1$ correspond to a light-like $V_\mu$, whereas for the circular polarization and bi-refringence, we have considered $V_\mu$ space-like.
We then conclude that, since we are dealing with different space-time classes
of $V_\mu$, triple refraction is not actually taking place.

 
\section{Wave energy loss}


\subsection{ CPT-odd classes}


In the CFJ scenario, we now study an EM excitation of a photon propagating in a constant external
field. The total field is given by

\begin{eqnarray}
{\vec{E}} & = & {\vec{E}}_{B}+{\vec{e}}~,\label{splittinge}\\
\vec{B} & = & \vec{B}_{B}+\vec{b}~,
\label{splittingb}
\end{eqnarray}
where $\vec{E}_{B}$ ($\vec{B}_{B}$) is the external electric (magnetic)
field. 
We first take the external field as uniform and constant, and thus

\begin{eqnarray}
\vec{\nabla}\cdot\vec{e}-\vec{V}\cdot\vec{b} & = & \Rho~,\\
\vec{\nabla}\times\vec{e} & = & -\partial_{t}\vec{b}~,\\
\vec{\nabla}\cdot\vec{b} & = & 0~,\\
\vec{\nabla}\times\vec{b}-V_{0}\vec{b}+\vec{V}\times\vec{e} & = & \partial_{t}\vec{e}+\vec{J}~,
\end{eqnarray}
where 
\beq
\Rho = \rho-\vec{V}\cdot\vec{B}_{B}~, 
\eeq
being $\rho$ the external charge density, and the other term the effective
charge due to the coupling between background and external field. Similarly $\vec{J}$ is the total current given by
\beq
\vec{J}=\vec{j}+V_{0}\vec{B}_{B}-\vec{V}\times\vec{E}_{B}~,
\eeq
in which $\vec{j}$ is the external current density and the other { terms are the effective
currents} due to the field coupling. From these equations, we get

\beq
\begin{cases}
\left(\vec{\nabla}\times\vec{e}\right)\cdot\vec{b}=-\partial_{t}\left(\frac{1}{2}\vec{b}^{2}\right)\\
\left(\vec{\nabla}\times\vec{b}\right)\cdot\vec{e}-V_{0}\vec{e}\cdot\vec{b}=\partial_{t}\left(\frac{1}{2}\vec{e}^{2}\right)+\vec{J}\cdot\vec{e}~.
\end{cases}
\eeq

Subtracting the first to the second, we obtain

\begin{gather}
\left(\vec{\nabla}\times\vec{b}\right)\cdot\vec{e}-\left(\vec{\nabla}\times\vec{e}\right)\cdot\vec{b}-V_{0}\vec{e}\cdot\vec{b}\nonumber\\
\qquad=\partial_{t}\left(\frac{1}{2}\vec{e}^{2}+\frac{1}{2}\vec{b}^{2}\right)+\vec{J}\cdot\vec{e}~.
\end{gather}

The first two terms can be rewritten as
\beq
\left(\vec{\nabla}\times\vec{b}\right)\cdot\vec{e}-\left(\vec{\nabla}\times\vec{e}\right)\cdot\vec{b}=\vec{\nabla}\cdot \left(\vec{b}\times\vec{e}\right)=-\vec{\nabla}\cdot\left(\vec{e}\times\vec{b}\right)~.
\eeq

Rewriting $\vec{e}\cdot\vec{b}$ as

\begin{equation}
\vec{e}\cdot\vec{b}=-\frac{1}{2}\partial_{t}\left(\vec{a}\cdot\vec{b}\right)+\vec{\nabla}\cdot\left(\frac{1}{2}\vec{a}\times\vec{e}-\frac{1}{2}\phi\vec{b}\right)~,
\end{equation} 

where $\vec{a}$ ($\phi$) is the magnetic (electric) potential of
the excitation, { it} yields { the non-conservation of the energy-moment\-um tensor} 

\begin{align}
& \vec{\nabla}\cdot\left(\vec{e}\times\vec{b}-V_{0}\phi\vec{b}+\frac{1}{2}V_{0}\vec{a}\times\dot{\vec{e}}\right)+ 
\partial_{t}\left(\frac{1}{2}\vec{e}^{2}+\frac{1}{2}\vec{b}^{2}-\frac{1}{2}V_{0}\vec{a}\cdot\vec{b}\right)\nonumber\\
& = -\left(\vec{j}+ V_{0}\vec{B}_{B}-\vec{V}\times\vec{E}_{B}\right)\cdot\vec{e}~.
\label{old134}
\end{align}

We observe that even when $\vec{j}=0$, there is dissipation, due to the coupling between the LSV background
and the external field. Thereby, in the CFJ scenario accompanied by an external field, the propagating wave $\left(\vec{e},\vec{b}\right)$
loses energy. 

Since in Eq. (\ref{old134}) the background vector $V_\mu$, and the external field, which is treated non-dynamically,
are both space-time-independent, they are not expected to contribute to the non-conservation of the energy-momentum tensor, for they do not introduce any explicit $x_{\mu}$ dependence in the CFJ Lagrangian, Eq. (\ref{CFJ_lagrangian}). However, there is here a subtlety. The LSV term, which is of the CS type, depends on the four-potential, $A_\mu$. By introducing the constant external fields, $E_B$ and  $B_B$, and performing the splittings of Eqs. ({\ref{splittinge},\ref{splittingb}), an explicit dependence on the background potentials, $\phi_B$  and  ${\vec A}_B$, appear now in the Lagrangian. 
But, if the background fields are constant, the background potentials must necessarily display
linear dependence on $x_\mu$ ($A^\mu_B = {\ds \frac{1}{2}}F_B^{\mu\nu}x_\nu$); the translation invariance of the Lagrangian is thereby lost. Then 
the LSV term triggers the appearance of the term $V_{0}\vec{B}_{B}-\vec{V}\times\vec{E}_{B}$ in the right-hand side of Eq. (\ref{old134}).

The above results may also be presented in the covariant formulation. We profit to include a non-constant external field in our setting, generalising the results above. On the other hand, we retain $V^{\mu}$ constant over space-time, to appreciate whether dissipation emerges with a minimal set of requirements on the LSV vector. We start off from
\begin{equation}
\partial_{\mu}F^{\mu\nu}+V_{\mu}\ ^{*}F^{\mu\nu}=j^{\nu}~,
\end{equation}
where $^{*}F^{\mu\nu}$ is the dual EM tensor field. We note the splitting
\begin{equation}
F^{\mu\nu}=F_{B}^{\mu\nu}+f^{\mu\nu}~,
\end{equation}
where $F_{B}^{\mu\nu}$ stands for the background electromagnetic field tensor and $f_{\mu\nu}$ corresponds to the 
propagating wave ($\vec{e}$,$\vec{b}$), both being $x_{\mu} $ dependent. We write the energy-momentum for the photon field ($f^{\mu\nu})$ as

\begin{equation}
\left(\theta_{f}\right)_{\ \rho}^{\mu}=f^{\mu\nu}f_{\nu\rho}+\frac{1}{4}\delta_{\rho}^{\mu}f^{2}-\frac{1}{2}\ ^{*}f^{\mu\nu}a_{\nu}v_{\rho}~,
\label{em1}
\end{equation}
where $^{*}f^{\mu\nu}$ is the dual EM tensor photon field. The first two terms of Eq. (\ref{em1}) are Maxwellian, whereas the third originates from the CFJ model. The photon energy-momentum tensor continuity equation reads as 

\begin{align}
\partial_\mu \left(\theta_{f}\right)_{\ \rho}^{\mu}=&j^\mu f_{\mu\rho} - V_\mu F_B^{\mu\nu}f_{\nu\rho} - \left (\partial_{\mu}F_B^{\mu\nu}\right) f_{\nu\rho} \nonumber \\
& -\frac{1}{2}\ ^{*}f^{\mu\nu}a_{\nu}\partial_\mu v_{\rho}~.
\label{ce1}
\end{align}

Equation (\ref{em1}) shows that the energy-momentum tensor, in presence of LSV terms, 
is no longer symmetric, as it had been long ago pointed out \cite{cafija90,colladaykostelecky1998}. 
In this situation, $\theta^{00}$ describes the field energy density; $\theta^{i0}$ represents the components of a generalised Poynting vector, while $\theta^{0i}$ is the 
true field momentum density.

If we denote the energy density by $u$ and the generalised Poynting vector as
$\vec{S}$, it follows that
\begin{align}
\partial_{t}u+\vec{\nabla}\cdot\vec{S} & =-\vec{j}\cdot\vec{e}-V^{0}\vec{E}_{B}\cdot\vec{e}-\left(\partial_{t}\vec{E}_{B}\right)\cdot\vec{e}+ \nonumber \\
 & \ \ \left(\vec{\nabla}\times\vec{B}_{B}\right)\cdot\vec{e}-\left(\vec{B}_{B}\times\vec{e}\right)\cdot\vec{V}~.
\label{finenoddloss}
\end{align}

Besides the external current $j^\mu$, external electric and magnetic fields (space-time constant
or not) are sources for the exchange of energy with the propagating ${\vec e}-$ and ${\vec b}-$ waves. In the
special case the external ${\vec E_B}-$ and ${\vec B_B}-$ fields are constant over space-time, their coupling to the components of the LSV vector are still responsible for the energy exchange with the electromagnetic
signals.} 


\subsection{CPT-odd and CPT-even classes}


Let us consider the field equation with both $V_\mu$ and $k_{F}$ space-time dependent; the Lagrangians Eqs. (\ref{CFJ_lagrangian},\ref{L3}) yield the field equations 

\beq
\partial_{\mu}F^{\mu\nu}+V_{\mu}\ ^{*}F^{\mu\nu}+\left(\partial_{\mu}k_{F}^{\mu\nu\kappa\lambda}\right)F_{\kappa\lambda}
+k_{F}^{\mu\nu\kappa\lambda}\partial_{\mu}F_{\kappa\lambda} 
=j^{\nu}~.
\eeq

We perform the same splitting as above
\begin{equation}
F_{\mu\nu}=\left(F_{B}\right)_{\mu\nu}+f_{\mu\nu}\ .
\end{equation}

We compute the energy-momentum tensor $\theta_{\ \rho}^{\mu}$ and its conservation
equation for the propagating signal $f_{\mu\nu}$

\begin{align}
\theta_{\ \rho}^{\mu} = &  
f^{\mu\nu}f_{\nu\rho} 
+ \frac{1}{4}\delta_{\rho}^{\mu}f^{2} 
- \frac{1}{2} {^*\!}f^{\mu\nu} a_\nu V_\rho \nonumber \\ 
& + k_{F}^{\mu\nu\kappa\lambda}f_{\kappa\lambda}f_{\nu\rho}
+ \frac{1}{4}\delta_{\rho}^{\mu}k_{F}^{\kappa\lambda\alpha\beta}f_{\kappa\lambda}f_{\alpha\beta}~,
\label{tempair}
\end{align}
and
\begin{align}
\partial_{\mu}\theta_{\ \rho}^{\mu} = & 
j^\nu f_{\nu\rho} 
- \left ( \partial_{\rho}F_{B}^{\mu\nu}\right)f_{\nu\rho}
- V_\mu {^*\!}F_{B}^{\mu\nu}f_{\nu\rho}\nonumber \\
& - \frac{1}{2} \left ( \partial_{\mu}V_\rho \right) {^*\!}f^{\mu\nu}a_{\nu}
+\frac{1}{4} \left(\partial_\rho k_{F}^{\mu\nu\kappa\lambda} \right) f_{\mu\nu}f_{\kappa\lambda}\nonumber \\
& - \left(\partial_\mu k_{F}^{\mu\nu\kappa\lambda} \right) F_{B\kappa\lambda}f_{\nu\rho}
- k_{F}^{\mu\nu\kappa\lambda} \left(\partial_\mu F_{B\kappa\lambda} \right) f_{\nu\rho}~.
\label{constempair}
\end{align}

The conservation equation { of the energy-momentum} cor\-responds to taking the
$\theta_{\ 0}^{\mu}$ component of the continuity equation, Eq. (\ref{constempair}).

The background time derivative terms $\left(\partial_{t}F_{B}^{\mu\nu}\right)f_{\mu\nu}$
and $k_{F}^{\mu\nu\kappa\lambda}\left(\partial_{t}F_{B\kappa\lambda}\right)f_{\mu\nu}$
may account for a deviation from the conservation of the energy-momentum
tensor of the propagating wave, whenever one of the fields $\vec{E}_{B}$, $\vec{B}_{B}$ is not constant.
 

\subsubsection{Varying breaking vector $V_{\mu}$ and tensor $k_F$ without an external EM field}


We deal with both CPT sectors at once. Indeed,  we start off from the Lagrangian  

\beq
L=-\frac{1}{4}\left(F^{\mu\nu}\right)^{2}+\frac{1}{4}\epsilon^{\mu\nu\kappa\lambda}V_{\mu}A_{\nu}F_{\kappa\lambda}-  
\frac{1}{4}(k_{F})_{\mu\nu\kappa\lambda}F^{\mu\nu}F^{\kappa\lambda}~,
\eeq
with $V_{\mu}$ and $k_{F}$ both $x^{\mu}$ dependent, and $n_{\mu}$ a constant four-vector. This Lagrangian is a combination of 
contributions from the breaking terms $V_\mu$ and $k_F$.
The resulting field equation is 

\beq
\partial_{\mu}F^{\mu\nu}+V_{\mu}{}{^*\!}{F}^{\mu\nu}+\partial_{\mu}[(k_{F})^{\kappa\lambda\mu\nu}F_{\kappa\lambda}]
=0~.  
\label{fe}  
\eeq

From Eq. (\ref{fe}), the  equation on energy-momentum follows

\begin{gather*}
\theta^{\mu}_{~\rho}=F^{\mu\nu}F_{\nu\rho}+\frac{1}{4}\delta^{\mu}_{\rho}F^{2}-\frac{1}{2}({}{^*\!}{F}^{\mu\alpha}
A_{\alpha}V_{\rho}) + \\
(k_{F})^{\kappa\lambda\mu\nu}F_{\kappa\lambda}F_{\nu\rho}+\frac{1}{4}\delta^{\mu}_{\rho}(k_{F})^{\kappa\lambda\alpha\beta}F_{\kappa\lambda}F_{\alpha\beta}~, 
\end{gather*} 
as well as its non-conservation

\beq
\partial_{\mu}\theta^{\mu}_{~\nu}=-\frac{1}{2}(\partial_{\mu}V_{\nu}){}{^*\!}{F}^{\mu\rho}A_{\rho}+\frac{1}{4}
\left(\partial_{\nu}k_{F}^{\mu\rho\kappa\lambda}\right)F_{\mu\rho}F_{\kappa\lambda}~.
\label{tiredwave}
\eeq

Equation (\ref{tiredwave}) confirms { that}, if $V_{\mu}$ and $k_{F}$ are coordinate dependent, there is 
energy and momentum exchange, and thereby dissipation even in absence of an external EM field. 
The LSV background  introduces an explicit space-time dependency in the Lagrangian so that the energy and momentum of the propagating electromagnetic field are not conserved.

If we take the energy density $\theta_{\ 0}^{0}:=u$ and the { generalised Poynting
vector $\theta^{0i}=\vec{S}$, we write}, from Eq. (\ref{tiredwave})

\begin{eqnarray}
\partial_{t}u+\vec{\nabla}\cdot\vec{S} & = & -\frac{1}{2}\left(\partial_{t}V_{0}\right)\vec{E}\cdot\vec{A}-\frac{1}{2}\left(\vec{\nabla}V_{0}\cdot\vec{B}\right)\Phi - \label{New continuity}\\
 &  & \frac{1}{2}\left(\vec{\nabla}V_{0}\times\vec{E}\right)\cdot\vec{A}+\frac{1}{4}\left(\partial_{t}k_{F}^{\mu\rho\kappa\lambda}\right)F_{\mu\rho}F_{\kappa\lambda}\ .\nonumber 
\end{eqnarray}

Therefore, it becomes clear that the CPT-odd term contributes to the
breaking of the energy-momentum conservation through the $V_{0}$
component; on the other hand, the CPT-even $k_{F}$ tensor affects
the energy continuity equation only if its components exhibit
time dependency. If $k_{F}^{\mu\rho\kappa\lambda}$ are only space
dependent, then there is no contribution to the right-hand side of
Eq. (\ref{New continuity}). 

Recalling that $\theta^{\mu\nu}$  is no longer symmetric in presence of a LSV background, if we consider the continuity equation for the momentum density of the field, described by $\theta^{0i}$, it can be readily checked that the space component of  $V_\mu$, ${\vec V}$, through its space and time dependencies, and the space dependency of the $k_F$ components will be also responsible for the non-conservation of the momentum density carried by the electromagnetic signals.


\subsubsection{The most general situation: LSV background and external field $x_{\mu}$-dependent}


In this Section, we present the most general case to describe the energy-momentum continuity equation for the photon
field ($f^{\mu\nu}$). 
By starting off from the field equation

\beq
\partial_{\mu}F^{\mu\nu}+V_{\mu}\ ^{*}F^{\mu\nu}+\left(\partial_{\mu}k_{F}^{\mu\nu\kappa\lambda}\right)F_{\kappa\lambda}
+k_{F}^{\mu\nu\kappa\lambda}\partial_{\mu}F_{\kappa\lambda} 
=j^{\nu}~,
\eeq
and using 
\beq
 ^{*}f^{\mu\rho}f_{\rho\nu}  =  -\frac{1}{4}\delta_{\nu}^{\mu}\ ^{*}f\cdot f =  -\frac{1}{2}\delta_{\nu}^{\mu}\partial_{\rho}\left(^{*}f^{\rho\lambda}a_{\lambda}\right)~,
\eeq
\beq
\left(\partial_{\mu}{}^{*}f^{\kappa\lambda}\right)f_{\kappa\lambda} =^{*}f^{\kappa\lambda}\left(\partial_{\mu}f_{\kappa\lambda}\right)=\partial_{\mu}\partial_{\kappa}\left(^{*}f^{\kappa\lambda}a_{\lambda}\right)~, 
\eeq

\begin{align}
k_{F}^{\mu\nu\kappa\lambda}f_{\kappa\lambda}\partial_{\nu}f_{\mu\rho} = & -\partial_{\rho}\left(\frac{1}{4}k_{F}^{\mu\nu\kappa\lambda}f_{\mu\nu}f_{\kappa\lambda}\right) + \nonumber \\
& \frac{1}{4}\left(\partial_{\rho}k_{F}^{\mu\nu\kappa\lambda}\right)f_{\mu\nu}f_{\kappa\lambda}\ ,
\end{align}
we present the photon energy-momentum tensor 

\begin{align}
\theta_{\ \rho}^{\mu} & =f^{\mu\nu}f_{\nu\rho}+\frac{1}{4}\delta_{\rho}^{\mu}f^{2}-\frac{1}{2}V_{\rho}\ ^{*}f^{\mu\nu}a_{\nu}+\nonumber \\
 & \ \ k_{F}^{\mu\nu\kappa\lambda}f_{\kappa\lambda}f_{\nu\rho}+\frac{1}{4}\delta_{\rho}^{\mu}k_{F}^{\kappa\lambda\alpha\beta}f_{\kappa\lambda}f_{\alpha\beta}~,
\end{align}
and its non-conservation
\begin{align}
\partial_{\mu}\theta_{\ \rho}^{\mu} & =j^{\nu}f_{\nu\rho}-\left(\partial_{\mu}F_{B}^{\mu\nu}\right)f_{\nu\rho}-V_{\mu}\ ^{*}F_{B}^{\mu\nu}f_{\nu\rho} - \nonumber \\
 & \ \ \frac{1}{2}\left(\partial_{\mu}V_{\rho}\right)\ ^{*}f^{\mu\nu}a_{\nu}+\frac{1}{4}\left(\partial_{\rho}k_{F}^{\mu\nu\kappa\lambda}\right)f_{\mu\nu}f_{\kappa\lambda} - \nonumber \\
 & \ \ \left(\partial_{\mu}k_{F}^{\mu\nu\kappa\lambda}\right)F_{B\kappa\lambda}f_{\nu\rho}-k_{F}^{\mu\nu\kappa\lambda}\left(\partial_{\mu}F_{B\kappa\lambda}\right)f_{\nu\rho}~.
\label{finalthetap}
\end{align}

The right hand-side of Eq.(\ref{finalthetap}) displays all
types of terms that describe the exchange of energy between the photon,
the LSV background and the external field, taking into account an { $x^\mu$-dependence} of the LSV background and the external field. 

In Eq. (\ref{finalthetap}), the first two right-hand side terms are purely Maxwellian. Further, since $\theta^\mu_{~~\nu}$ is not symmetric in presence of LSV terms, when taking its four-divergence with respect to its second index, namely 
$\partial^\nu \theta^\mu_{~~\nu}$, contributions of the forms $\partial^\nu k_{F\kappa\lambda\nu\rho}F^{\kappa\lambda}f^{\rho\mu}$ and 
$\partial^\nu k_{F}^{\kappa\lambda\mu\rho}F_{\kappa\lambda}f_{\rho\nu}$ appear. Thus, even when $k_{F}^{\kappa\lambda\mu\rho}$ is only space dependent, though not contributing to $\partial_\nu \theta^{\nu 0}$, it does contribute to $\partial_\nu \theta^{0\nu}$.  
We observe that the roles of the perturbation vector and tensor differ, the latter demanding a space-time dependence of the tensor or of the external field, conversely to the former.   

As final remark, the energy losses would presumably translate into frequency damping if
the excitation were a photon. Whether such losses could be perceived as 'tired light' needs an analysis of the wave-particle relation.  


\section{Conclusions, discussion and perspectives}


We have approached the question of non-Maxwellian photons from a more fundamental perspective, linking their
appearance to the breaking of the Lorentz symmetry. Despite massive photons have been proposed in several works, few  hypothesis on the mass origin have been published, see for instance \cite{addvgr07}, and surely there is no comprehensive discussion taking form of a review on such origin, see for instance \cite{goni10}. It is our belief that answering
this question is a crucial task in order to truly understand the nature of
the electromagnetic interaction carrier and the potential implications in interpreting signals from the Universe. Given the complexity of the subject, we intend to carry on our research in future works.

The chosen approach concerns well established SuSy theories that go beyond the Standard Model. 
Some models originated from SuSy\footnote{Other models are outside SuSy. Identical results are found in \cite{adamklinkhamer2001,bsbebohn2003}.}: see for instance \cite{baetaetal2004,kome02,bebegahnle2015} determined dispersion relations, but the  
analysis of the latter was unachieved. We also derived the dispersion relations for those cases not present in the literature and also for those we 
charged ourselves with the task of studying the consequences in some detail. We did not intend to cover all physical cases, and 
we do not have any pretense of having done so. Nevertheless, we have explored quite a range of both odd and even CPT sectors.
 
We stand on the conviction that a fundamental theory describing nature should include both CPT sectors. The understanding of the interaction between the two sectors is far from being unfolded and one 
major question remains open. If we are confronted with a non-Maxwellian behaviour for one sector, or worse for two sectors, how would a two-sector theory narrate the propagation? Would the two contributions be simply additive or would there be more interwoven relations?
The answers to these questions would prompt other stimulating future avenues of research.

Starting from the actions representing odd and even CPT sector, for both we have analysed whether the photon propagation is impacted by its SuSy partner, the photino.
Though the SuSy partners have not been experimentally detected yet, it is possible to assess their impact. Indeed, the actions of Eqs. (\ref{Lagrangian2},\ref{Lclass4}), describe effective photonic models for which the effects of the photino have been summed up at the classical level, that is without loop corrections. Thus, the corresponding DRs include SuSy through the background of the Fermionic sector accompanying the $V^\mu$ and $k_F$ breaking vector and tensor, respectively. It would be worth to draw from the constraints on the SME coefficients the estimates of the background SuSy condensates. The latter when related to the SuSy breaking scale and thereby to the masses of the SuSy partners, and specifically the photino. This is a relevant issue for investigating the connection between the SuSy breaking scale, associated to the condensates of the Fermionic partners in the LSV background, and the constraints on the SME.

For the CPT-odd case, we study the super-symmetrised  \cite{baetaetal2004,bebegahnle2015} Carroll-Field-Jackiw
model \cite{cafija90},  where the Lorentz-Poincar\'e symmetry violation is determined by the $V_\mu$ four-vector. The resulting dispersion relation is of the fourth order. 

For the next conclusions, we do not distinguish between classes with respect to photino integration. 

In short, the major findings can be summarised as follows. 
For the effective photon mass:
\begin{itemize}
\item{Whenever an explicit solution is determined, at least one solution shows a massive photon behaviour. It is characterised by a frequency dependency of the type $\omega^{-2}$ like the classic de Broglie-Proca photon.}
\item{The mass is effective and proportional to the absolute value of the Lorentz symmetry breaking vector. The gro\-und based upper limits \cite{gomesmalta2016} are compatible with state of the art experimental findings on photon mass \cite{tanabashietal2018}.} 
\item{The group velocity is almost always sub-luminal. Super-luminal speeds may appear if the time component of the breaking vector differs from zero. They appear beyond a frequency threshold.}  
\item{The photon mass is gauge invariant as drawn by the Carr\-oll-Field-Jackiw model, conversely to the de Broglie-Proca photon.}  
\item{Bi-refringence accompanies the CPT-odd sector.}
\end{itemize}

Other notable features are

\begin{itemize}
\item{When the time component of the LSV breaking vector differs from zero, imaginary and complex frequencies may arise.}
\item{ We have determined group velocities in the following cases: when the time component or the along the line of sight component of the breaking vector vanishes. The most general case, all components being present, was analysed for $V^\mu$ light-like.}
\item{The solutions feature anisotropy and lack of Lorentz invariance, due to the dependency on the angle between the breaking vector and the propagation direction, or else on the chosen reference frame.  }  
\item {Since two group velocities for the CPT-odd handedness were found except for $V^\mu$ light-like, we pursued an analysis of the dispersion relation in terms of the fields, in well defined polarisations. We have determined the existence of bi-refringence. }
\end{itemize}

Having recorded for almost all CPT-odd cases, a mass\-ive-like behaviour, we have explained this phenomenology tracing its origin back to  the Carroll-Field-Jackiw Lagrangian.  We have recast it in a non-explicit but still covariant form, introducing the photon field components.
The electric potential is not a dynamical variable and we eliminated it from the Lagrangian. In the latter, a term that
has the classic structure of the de Broglie-Proca photon mass arises, where the breaking vector playing the role of the mass. This
is consistent with what we had previously seen in the dispersion relations. It gives us a more fundamental reason
for which the mass of the photon would be linked to the breaking vector.

For the CPT-even sector, we adopt the $k_{F}$ breaking tensor model \cite{bebegahnle2015}. From the dispersion relations, we evince 

\begin{itemize}
\item{Generally, being the propagation of the photon affected by the action of the breaking tensor, we have a tensorial anisotropy and thereby a patent lack of Lorentz invariance. The main consequence is that the speed of light depends on the direction. The correction goes like the breaking components squared. As the components are tiny, since they represent the deviation from the 
Lorentz invariance, also the correction to $c$ will be limited to small values.}  
\item{Nevertheless, if the breaking tensor is proportional
to the Kroeneker's delta, the dispersion relation looks as a light ray propagating through a medium. The vacuum assumes an effective refraction index due to the interaction of the photon with the background.}
\item{ From the Class 3 Lagrangian, it follows that no mass can be generated for the photon. Indeed, the dispersion relation yields $\omega = 0$ whenever  ${\vec k} = 0$. Instead, for Class 4, there may take place a photon mass generation, due to the b-term which represents higher derivatives in the Lagrangian. Thus, the DR includes the possibility of a non-trivial $\omega$-solution even if we take a
trivial wave vector. }
\end{itemize}

Possibly, the most remarkable result concerns energy dissipation for both odd and even CPT sectors. 

\begin{itemize}
\item{In the odd sector, the coupling of a constant external field, with a constant breaking vector, determines an energy loss even in absence of an external current.  This is revealed by the breaking of the continuity equation
(or conservation) of the photon energy-momentum tensor. If the photon 
is coupled to the LSV background and/or an EM external field which explicitly
depend on the space-time coordinates, then translational symmetry
is broken and the energy-momentum tensor is no longer conserved.
This means that the system under consideration is exchanging energy
(loosing or even receiving) with the environment.}
\item{Still in the odd sector, in absence of an external field, but in presence of a space and/or time dependency of the time component of the breaking vector, energy loss occurs. }
\item {Finally, we have considered odd and even CPT sectors together. We found if $V_\mu$ and $k_F$ are coordinate dependent, there is dissipation in absence of an external EM field.  }
\end{itemize}

The relation between dissipation and complex, or simply imaginary, frequencies naturally arises. Perspectives in research stem from the issues below.

Dissipation occurs in both odd and even CPT sectors when the associated breaking factors are not constant over space-time (for the following considerations, we neglect any external field). 
However, in the odd sector, even if $V_\mu$ is
constant, complex frequencies may arise since the dispersion relation is quartic in frequency. This is due to the Carroll-Field-Jackiw model which does not ensure a positive-definite energy, and thereby we may have unstable
configurations. This leads to complex frequencies.
Imaginary frequencies imply damping which is associated to dissipation, and we don't feel having cleared the issue sufficiently.

The CPT-even sector does not get in trouble with the positiveness
of the energy, and thereby complex frequencies associated to unstable excitations
are absent. So, the CPT-even sector may yield dissipation, when $k_F$ is non-constant, even if it does not exhibit complex frequencies.

 In short, future analysis of dissipation will have to tackle and possibly set boundaries towards imaginary frequencies and 
super-luminal velocities, knowing that dissipation might very well occur for sub-luminal propagation.   

We shall be analysing these and related issues, in connection with the conjectures of tired light in forthcoming works, also in the frame of a classic non-linear formulation of electromagnetism. 
We take note of different but otherwise possibly converging efforts \cite{partanenetal2017}. 


\appendix


\section{On CPT-even classes\label{appeven}}


We intend to write the $k_F$ tensor in terms of a single Bosonic vector $\xi_{\mu}$ which signals LSV. This field is supposed
to be part of a chiral field of which the Fermionic condensates generate the LSV. For achieving this purpose, we start by neglecting the fully anti-symmetric part in $\left(k_{F}\right)_{\mu\nu\alpha\beta}$, since it would only account to a total derivative in the action (we exclude the component yielding bi-refringence, in this manner). Exploiting the {\it Ansatz} 
in \cite{bekakl2009,baileykostelecky2004}, we write for 

\beq
{\kappa}_{\alpha\beta}=\xi_{\alpha}\xi_{\beta}-\eta_{\alpha\beta}\frac{\xi_{\rho}\xi^{\rho}}{4}~, 
\label{a1}
\eeq
we have

\begin{eqnarray}
\left(k_{F}\right)_{\mu\nu\alpha\beta}&=&-\left(k_{F}\right)_{\nu\mu\alpha\beta}=-\left(k_{F}\right)_{\mu\nu\beta\alpha}=\left(k_{F}\right)_{\alpha\beta\mu\nu}\nonumber \\
&=&\frac{1}{2}\left(\eta_{\mu\alpha}{\kappa}_{\nu\beta}-\eta_{\mu\beta}{\kappa}_{\nu\alpha}+\eta_{\nu\beta}{\kappa}_{\mu\alpha}-\eta_{\nu\alpha}{\kappa}_{\mu\beta}\right)\nonumber\\
&:=& K_{\mu\nu\alpha\beta}~,\label{kF}\\
\bar{K}_{\alpha\beta} & = & K_{\mu\nu\alpha\beta}{ \bar{k}_{\mu}\bar{k}_{\nu}}~,\\
{ \bar{k}^{\mu}} & = & \frac{k^{\mu}}{|\vec{k}|}~.
\end{eqnarray}

This, in turn, implies a Lagrangian in the form

\beq
L_{3}=\frac{1}{4}\left(\frac{1}{2}\xi_{\mu}\xi_{\nu}F_{\alpha}^{\mu}F^{\alpha\nu}+\frac{1}{8}\xi_{\rho}\xi^{\rho}F_{\mu\nu}F^{\mu\nu}\right)~.\label{Lclass3}
\eeq

These simplifications are legitimate. In fact, had we taken into account the full complexity of the
$k_{F}$ term, then we would have had to deal with a higher spin super-field. Its appearance is instead avoided thanks to
transferring the effects to the $\xi^{\mu}$ vector. 

The Lagrangian in Eq. (\ref{Lclass4}) is obtained carrying out the
super-symmetrisation of Eq. (\ref{Lclass3}) taking into account that $\xi^{\mu}$ defines the SuSy breaking field.

We are interested in obtaining an effective photonic Lagrangian by integrating out the photino sector (and all others
SuSy sectors as well).
The resulting Lagrangian reads \cite{bebegahnle2015} as Eq. (\ref{Lclass4}). Since the DR for this theory is not present in literature, we proceed to its derivation. The steps are as usual the following: i) write the Lagrangian in terms of the fields; ii) get the Euler-Lagrange equations; iii) perform the Fourier transform. 

The Lagrangian in terms of the potential is \cite{bebegahnle2015}
 
\begin{eqnarray}
& L_4 & = \frac{1}{2}A^{\mu}\left[\left(\square - r \chi^{\alpha\beta}\partial_{\alpha}\partial_{\beta}\right)\eta_{\mu\nu} - 
\left(\partial_\nu - r \chi_{\nu\alpha}\partial^\alpha \right)\partial_\mu +
\right. \nonumber \\
& & r \chi_{\mu\alpha}\partial^{\alpha}\partial_{\nu} + \chi_{\mu\nu}\square(- r + s\square) 
- s (\chi_\nu^{~\alpha}\partial_\mu + \chi_\mu^{~\alpha}\partial_\nu ) \square \partial_\alpha  + \nonumber \\
 &  & \left. s \chi_{\alpha\beta}\partial^{\alpha}\partial^{\beta}\partial_{\mu}\partial_{\nu}\right ] A^{\nu}~.
\end{eqnarray}

Varying with respect to $A^{\mu}$ and
performing the Fourier transform, we obtain

\begin{eqnarray}
 & & \left[k^{2}\delta_{\mu}^{\nu} - r \chi^{\alpha\beta}k_{\alpha}k_{\beta} + r \chi_{\mu\alpha}k^{\alpha}k^{\nu}-\chi^{\nu}_{~\mu}k^2 
(r + s k^2)+\right.\nonumber\\
& &\left. s \chi_{\mu}^{~\alpha}k^{2}k_{\alpha}k^{\mu}\right]\tilde{A}^{\mu}=0~. 
\end{eqnarray}

having chosen the Lorenz gauge 

\beq
k_{\nu}\tilde {A}^{\nu}=0~.
\eeq

This shows that we have a matricial equation in the form

\beq
M_{\mu\nu}\tilde{A}^{\nu}=0~,
\eeq
which has non-trivial solutions only if 

\beq
\mbox{det}M_{\mu\nu}=0~.
\eeq

By rearranging the terms, we see that 

\begin{gather}
M^{\nu}_{~\mu}=k^{2}\left[\delta^{\nu}_{\mu}- r \left(\chi^{\alpha\beta}\frac{k_{\alpha}k_{\beta}}{k^{2}}\delta^{\nu}_{\mu}
- \chi_{\mu\alpha}\frac{k^{\alpha}k^{\nu}}{k^{2}} \right) - \right.  \nonumber \\
\left. \chi^\nu_{~\mu} ( r + s k^2) + s \chi_{\mu}^{~\alpha}k_\alpha k^\nu \right]
\label{Mmatrix}
\end{gather}
has the structure of the identity plus something small, since the parameters $r$ and $s$ are dependent upon the symmetries
violating terms which are extremely small. Therefore

\beq
\mbox{det}M_{\mu}^{\nu}=\mbox{det}\left(\mathbb{I}+X\right)=e^{\mbox{tr}\left[\mbox{ln}\left(\mathbb{I}+X\right)\right]}~,
\eeq
with $X$ small. Expanding the logarithm, 

\begin{eqnarray}
\mbox{det}\left(\mathbb{I}+X\right) & \sim & e^{\mbox{tr}\left[X-\frac{X^{2}}{2}\right]}\nonumber \\
 & = & e^{\mbox{tr}X-\frac{1}{2}\mbox{tr}X^{2}}\nonumber \\
 & \sim & 1+\mbox{tr}X-\frac{1}{2}\mbox{tr}X^{2}+\frac{1}{2}\left(\mbox{tr}X\right)^{2}+O\left(X^{3}\right)~.
\end{eqnarray}

Using Eq. (\ref{Mmatrix}) we finally obtain, at first order

\begin{gather}
s \chi k^{4}- \left (1 - r \chi + s \chi^{\alpha\beta}k_{\alpha}k_{\beta}\right )k^2 + 3 r \chi^{\alpha\beta}k_{\alpha}k_{\beta} = 0~, 
\label{DR-A14}
\end{gather}
where $\chi=\chi_{\mu}^{~\mu} = \chi^0_{~0}+ \chi^i_{~i}$. 
If we consider $\chi^{00} = \chi^{0i} =0$, then 
$\chi = \chi^1_{~1} + \chi^2_{~2} + \chi^3_{~3} = \chi_1 + \chi_2 + \chi_3$. 
We point out here that Eq. (\ref{DR-A14}), taken with $r = 0, s = 2\eta^2$ and $X_{\mu\nu} = D_{\mu\nu}$  reproduces the DR given in Eq. (29) of \cite{casana-ferreira-lisboasantos-dossantos-schreck-2018}, once the latter is linearised in the tensor $D_{\mu\nu}$ and taken with $\theta = 0$.


\section*{ Acknowledgements} 


LB and ADAMS acknowledge CBPF for hospitality, while LRdSF and JAHN are grateful to CNPq-Brasil for financial support. All authors thank the referee for the very detailed comments that improved the work. 

\bibliographystyle{unsrt}
\bibliography{references_spallicci_180831}

 \end{document}